\documentclass[12pt,preprint]{aastex}
\newcommand{\noprint}[1]{}

\shorttitle{GOALS NICMOS Imaging}
\shortauthors{Haan et al.}

\begin{document}

\title{The Nuclear Structure in Nearby Luminous Infrared Galaxies: HST NICMOS Imaging of the GOALS Sample}

\author{S. Haan\altaffilmark{1}, J.A. Surace\altaffilmark{1}, L. Armus\altaffilmark{1}, A.S. Evans\altaffilmark{2,3}, J.H. Howell\altaffilmark{1}, J.M. Mazzarella\altaffilmark{4}, D.C. Kim\altaffilmark{2}, T. Vavilkin\altaffilmark{5}, H. Inami\altaffilmark{1,6}, D.B. Sanders\altaffilmark{7}, A. Petric\altaffilmark{1}, C.R. Bridge\altaffilmark{8}, J.L. Melbourne\altaffilmark{8}, V. Charmandaris\altaffilmark{9,10}, T. Diaz-Santos\altaffilmark{9}, E.J. Murphy\altaffilmark{1}, V. U\altaffilmark{7}, S. Stierwalt\altaffilmark{1},  J.A. Marshall\altaffilmark{11}}

\altaffiltext{1}{Spitzer Science Center, California Institute of Technology, Pasadena, CA 91125, USA}
\altaffiltext{2}{National Radio Astronomy Observatory, Charlottesville, VA 22903, USA}
\altaffiltext{3}{Department of Astronomy, University of Virginia, Charlottesville, VA 22904, USA}
\altaffiltext{4}{Infrared Processing and Analysis Center, California Institute of Technology, Pasadena, CA 91125, USA}
\altaffiltext{5}{Department of Physics and Astronomy, SUNY Stony Brook, Stony Brook, NY 11794, USA}
\altaffiltext{6}{Department of Space and Astronautical Science, The Graduate University for Advanced Studies, Japan}
\altaffiltext{7}{Institute for Astronomy, University of Hawaii, Honolulu, HI 96822, USA}
\altaffiltext{8}{California Institute of Technology, Pasadena, CA 91125, USA}
\altaffiltext{9}{Department of Physics and Institute of Theoretical and Computational Physics, University of Crete, GR-71003, Heraklion}
\altaffiltext{10}{IESL/Foundation for Research and Technology - Hellas, GR-71110, Heraklion, Greece and Chercheur Associ\'{e}, Observatoire de Paris, F-75014, Paris, France}
\altaffiltext{11}{The Jet Propulsion Laboratory, California Institute of Technology, Pasadena, CA 91125, USA}

\begin{abstract}
We present results of Hubble Space Telescope NICMOS H-band imaging of 73 of most luminous (i.e., log[$L_{IR}/L_{\odot}]>11.4$) Infrared Galaxies (LIRGs) in the Great Observatories All-sky LIRG Survey (GOALS). This dataset combines multi-wavelength imaging and spectroscopic data from space (Spitzer, HST, GALEX, and Chandra) and ground-based telescopes. In this paper we use the high-resolution near-infrared data to recover nuclear structure that is obscured by dust at optical wavelengths and measure the evolution in this structure along the merger sequence. A large fraction of all galaxies in our sample possess double nuclei ($\sim$63\%) or show evidence for triple nuclei ($\sim$6\%). Half of these double nuclei are not visible in the HST B-band images due to dust obscuration.
The majority of interacting LIRGs have remaining merger timescales of 0.3 to 1.3 Gyrs, based on the projected nuclear separations and the mass ratio of nuclei. We find that the bulge luminosity surface density $\mathrm{L_{Bulge}/R_{Bulge}^2}$ increases significantly along the merger sequence (primarily due to a decrease of the bulge radius), while the bulge luminosity shows a small increase towards late merger stages. No significant increase of the bulge S\'{e}rsic index is found. LIRGs that show no interaction features have on average a significantly larger bulge luminosity, suggesting that non merging LIRGs have larger bulge masses than merging LIRGs. This may be related to the flux limited nature of the sample and the fact that mergers can significantly boost the IR luminosity of otherwise low luminosity galaxies. 
We find that the projected nuclear separation is significantly smaller for ULIRGs (median value of 1.2~kpc) than for LIRGs (mean value of 6.7~kpc), suggesting that the LIRG phase appears earlier in mergers than the ULIRG phase.
\end{abstract}


\section{Introduction}

Luminous IR Galaxies (LIRGs: i.e., $L_{IR} \geq 10^{11} L_{\odot}$) were discovered in the 1980s with the Infra Red Astronomical Satellite \citep[IRAS;][]{Soi84}, due to their strong emission at 12-100$\mu$m, and today are an important class of objects for understanding massive galaxy formation and evolution. Follow-up observations with ISO and Spitzer have shown that LIRGs comprise a significant fraction (50-70\%) of the cosmic infrared background and dominate the star-formation activity at $z\geq1$ \citep{Elb02, LeF05, Cap07, Bri07, Mag09}.  Ultra Luminous IR Galaxies (ULIRGs: i.e., $L_{IR} \geq 10^{12} L_{\odot}$) are typically a mixture of disk galaxy pairs, interactions, or mergers \citep{Jos85, Arm87, San88b, Cle96,Mur96}. (U)LIRGs are also observed to be rich in molecular gas and dust \citep{San91, Sol97}; interaction drives the gas inward, resulting in enhanced star formation rates. In the local universe, ULIRGs are relatively rare, but are about a thousand times more frequent at $z\geq 2$ as shown by observations of the population of sub-mm galaxies \citep{Bla02, Cha05}.\par 

Recent discoveries have revealed that the central black hole mass is constrained by and closely related to properties of the host galaxy's bulge \citep[e.g. ][]{Mar03,Har04, Geb00}, which suggests a co-evolution of the central black hole and the galaxy itself. Given both the evidence that merging LIRGs likely evolve into massive elliptical and S0 galaxies \citep[e.g. ][]{Bar92, Gen01, Tac02} and that LIRGs are stellar nurseries \citep[see e.g.][]{San91, Sol97, Mag09, How10}, the study of nuclear regions of a complete sample of LIRGs spanning all merger and interaction stages allows us to investigate the coeval nature of both processes.\par

The picture that mergers transform disc galaxies into elliptical galaxies is supported by numerical simulations \citep[e.g.][]{Bar90, Kor92}, potentially linking LIRG activity to a key step in the formation of the galaxy morphologies seen locally. Other recent simulations have shown evidence that this may be more complicated \citep[e.g.][]{Bou05, Rob06}, suggesting the importance of different merger-mass ratios, gas-fraction and feedback processes due to supernovae or AGN activity. Moreover, the exact number of merger progenitors is controversial; while the vast majority of interacting (U)LIRGs are likely two galaxy systems \citep[e.g.][]{Mur96,Vei02}, there may be also a small fraction of LIRGs that involve three galaxies and substantial star-formation between the merging nuclei \citep{Vai08}. While previous ground-based observations \citep[e.g.][]{Wri90, Jam99, Rot04} have revealed that a significant fraction of merger remnants seems to exhibit radial profiles similar to elliptical galaxies based on their large-scale appearance (at radii of several kpc), we are able to probe the first time the stellar light distribution in the central kpc for a large sample of nearby mergers. \par 

Merger models that incorporate hydrodynamics and star formation suggest that dissipation in mergers produces central starbursts which lead to a steepening of the central profile well above the extrapolation of a \cite{deV48} profile fit to the main body of the galaxy \citep{Her93, Mih94}. Indeed, recent observations have revealed such central ``extra light'' or ``cuspiness'' components for some elliptical galaxies \citep{Rot04, Fer06, Cot07, Kor09}. In particular, \cite{Hop08a} have shown that the observed extra light can be identified with the central density excess produced in simulations of gas-rich mergers and that such components are ubiquitous in the local cusp elliptical population \citep{Hop09a}, based on samples from \cite{Lau07} and \cite{Kor09}. However, although the idea that almost all the gas falls to the center to fuel a central starburst or AGN might be true for ULIRGs, this process may be less prevalent in LIRGs. In principle, a LIRG sample spanning all merger stages can be used to test the model of merger-induced cusp-building along the merger stage sequence. The Great Observatories All-sky LIRG Survey \citep[GOALS, see][]{Arm09} provides an excellent sample to compare to these models and observations. \par  

To reveal the detailed structure of the nuclear regions, where dust obscuration may mask star clusters, AGN and additional nuclei from optical view,  we study 73 of most luminous (U)LIRGs in the GOALS sample using the Hubble Space Telescope (HST) Near Infrared Camera and Multi-Object Spectrometer (NICMOS).  This high-resolution Near Infra-Red (NIR) imaging sample is unique not only in its completeness and sample size, but also in the proximity and brightness of the galaxies. Moreover, a wealth of additional information about the properties of these galaxies (e.g. mid-IR distribution, mid-IR line diagnostics, merger stage) is available, since the GOALS project combines multi-wavelength imaging and spectroscopic data from space (Spitzer, HST, GALEX, and Chandra) and ground-based telescopes. In particular, a detailed multi-wavelength study of the LIRG Markarian 266 \citep{Maz10} reveals nuclear and galactic-scale outflows, shock-heated gas and the presence of a dual AGN. \cite{Dia10} found that at least half of the MIR emission is extended for more than 30\% of local LIRGs, as well as an increasing compactness towards the final stage of major merger interaction. Furthermore, the relationship between the IR and UV properties \citep[e.g. the IRX-$\beta$ relation as a function of LIRG properties][]{How10} has been investigated. The optical GOALS HST images (ACS/WFC F435W and F814W filters) have clearly shown the inner spiral structure, dust lanes, extended filamentary emission, and star clusters in the nuclear regions and tidal tails of LIRGs \citep{Vav10, Kim10}.\par 

To understand the manner in which clusters, AGN, and nuclei evolve as a function of luminosity and merger stage, it is essential to first unveil the hidden nuclei and to model the nuclear structure. This requires both high spatial resolution ($\sim$100~pc) and the ability to penetrate the dust. Unfortunately, it is almost impossible to reveal the circumnuclear regions and to identify the nuclei in LIRGs with observations at optical wavelengths, simply because they are extremely dusty. On the other hand, NIR imaging is 1) less affected by dust extinction and hence provides an almost unobscured view on the nuclear region of galaxies, and 2) best suited to trace the old stellar population since NIR light is less biased by young blue stars that can dominate the optical light but contribute only a minor fraction by mass to the total stellar population. In particular, detailed information about the nuclear structure is crucial to reveal double or multiple nuclei to estimate merging time scales which can provide important constraints for the lifetime of the IR luminous phase and associated galaxy evolution scenarios. Therefore, HST NICMOS images are perfectly suited to observe more directly the true nuclear morphologies by combining the high resolution of the HST ($0.15\arcsec$ FWHM) which corresponds to 30 - 300~pc over the distance range 40 - 400~Mpc (with a median resolution of 106~pc at a median distance of 142~Mpc) at wavelengths where dust extinction is reduced by an order of magnitude compared to visual wavelengths.\par

Our NICMOS imaging has revealed the presence of double and triple nuclei LIRG systems. Fitting the NIR light distribution allows us to extract the structural stellar components of the galaxies and to study their evolution along the merger stage sequence. Our NICMOS sample and the observations are described in \S~\ref{sec:obs}. The modeling of the stellar components using GALFIT \citep{Pen02} and their parameters are characterized in \S~\ref{sec:galfit}, and the results are presented in \S~\ref{sec:res} together with the numbers of double and multiple nuclei found in our sample. In \S~\ref{sec:dis} we discuss the number of nuclei that are obscured by dust and its implication for high-redshift studies, merger time scales, the evolution of the central stellar morphology along the merger stage sequence, and the role of AGN activity and young stellar populations on nuclear properties.  
For convenience, we refer to the full sample (log[$L_{IR}/L_{\odot}]>11.4$) as LIRGs despite the inclusion of a small number (17 systems, $\sim$23\%) of ULIRGs (log[$L_{IR}/L_{\odot}]>12.0$).

\section{Observations and Data Reduction}
\label{sec:obs}
The GOALS NICMOS imaging program targets the nuclear regions of all systems in the IRAS Bright Galaxy Sample \citep[RGBS;][]{San03} with log[$L_{IR}/L_{\odot}]>11.4$. The RBGS is a complete sample of all extragalactic objects with $f_\nu(60\mu m) \geq 5.24$
Jy, covering the entire sky surveyed by IRAS at Galactic latitudes $\vert b\vert > 5^\circ$. The RBGS contains 203 LIRGs, of which 88 have luminosities of log[$L_{IR}/L_{\odot}]>11.4$, the luminosity at which the local space density of LIRGs exceeds that of optically-selected galaxies.
These galaxies are the most luminous members of the GOALS sample and they are predominantly mergers and strongly interacting galaxies. Most of the single spiral galaxies or widely separated, weakly/non- interacting pairs are filtered out by this luminosity threshold. The fraction of single spiral galaxies and non-interacting pairs in our NICMOS sample is $\sim$30\%.
The redshift range of our sample is $0.01 < z < 0.05$ and hence the galaxies are bright and well-resolved due to their large angular size. Here we present the first results of 73 (56 LIRGs, 17 ULIRGs) out of the 88 most luminous (U)LIRGs with a range in luminosity of (11.4$<$log[$L_{IR}/L_{\odot}]<12.5$) representing the high-luminosity end of the GOALS sample \citep{Arm09}.\par 

New HST images with NICMOS/NIC2 have been obtained using the F160W filter for 44 LIRGs within a total time of 59 orbits (program 11235, Surace P.I.). These data are combined with twenty-nine LIRGs that already had high-quality archival NICMOS data (see Tab.~\ref{tab:obs} for an overview of our sample). The data were collected using camera two (NIC2) with a field of view of $19.3\arcsec \times 19.5\arcsec$, a pixelsize of $0.075\arcsec$, and are dithered to yield a total field area of typically 30$\arcsec$ $\times$ 25$\arcsec$. Fifteen of the galaxies in our sample have two apparent optical nuclei separated by 15$\arcsec$--40$\arcsec$, or inclined disks with extensions greater than 20$\arcsec$, and thus required two pointings per orbit. Note that NICMOS failed during execution of this program, so that not all targets in the complete sample of 88 LIRGs were observed. The remainder are being observed with WF3, but these data are not yet collected. The data reduction and calibration has been done using the standard HST pipeline. Additional corrections were made to the individual frames to correct for bias offsets and we recombined the images using the STIS software. A World Coordinate System (WCS) correction could not be performed directly using cataloged star coordinates given the limited Field of View (FOV) of the NICMOS camera NIC2 ($19.2\arcsec \times 19.2\arcsec$). Instead the HST ACS images, which have a far larger FOV ($202\arcsec \times 202\arcsec$), have been calibrated with star position references (2MASS) in a first step and we subsequently carefully applied a WCS transformation of our NICMOS images given the corrected HST ACS images using corresponding reference points in both images (such as stars not cataloged, bright star cluster knots, or similar features).  Furthermore, GALFIT \citep{Pen02} is used to fit the F160W light from the underlying old stellar population and the AGN (see \S.~\ref{sec:galfit} for details).  
% –residuals from these fits will highlight structural features such as bars, funnels and bridges,and star clusters 

\section{Results}
\label{sec:res}

Here we present results of the nuclear stellar properties of 73 LIRG systems. Most of these systems are major merger, however, some LIRGs in our sample are isolated undisturbed galaxies, suggesting that other mechanisms rather than merger interaction may trigger the IR luminosity in those isolated undisturbed LIRGs (e.g. high gas fraction and stellar mass, cold gas accretion, or other internal dynamical processes). Therefore, these two sub-classes of LIRGs, interacting and isolated galaxies, are treated separately in our studies. Furthermore, to study the stellar nuclear properties as a function of merger stage (see \S~\ref{subsec:res-GALFIT}), we have applied a classification into different merger stages of our interacting LIRGs as described in the following.\par

\begin{figure}[h]
\begin{center}
\includegraphics[scale=0.42]{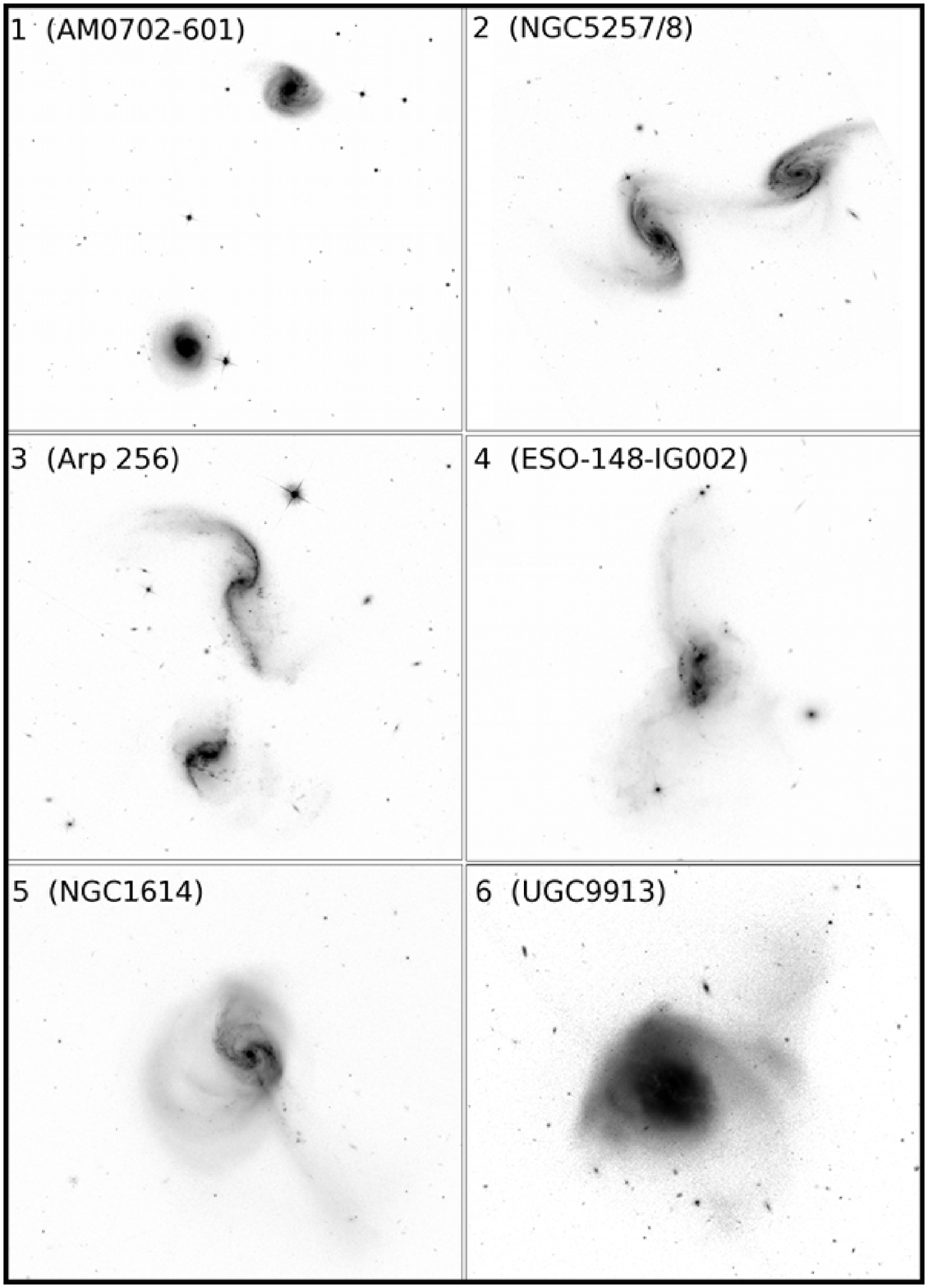}
\\
\medskip 
\includegraphics[scale=0.42]{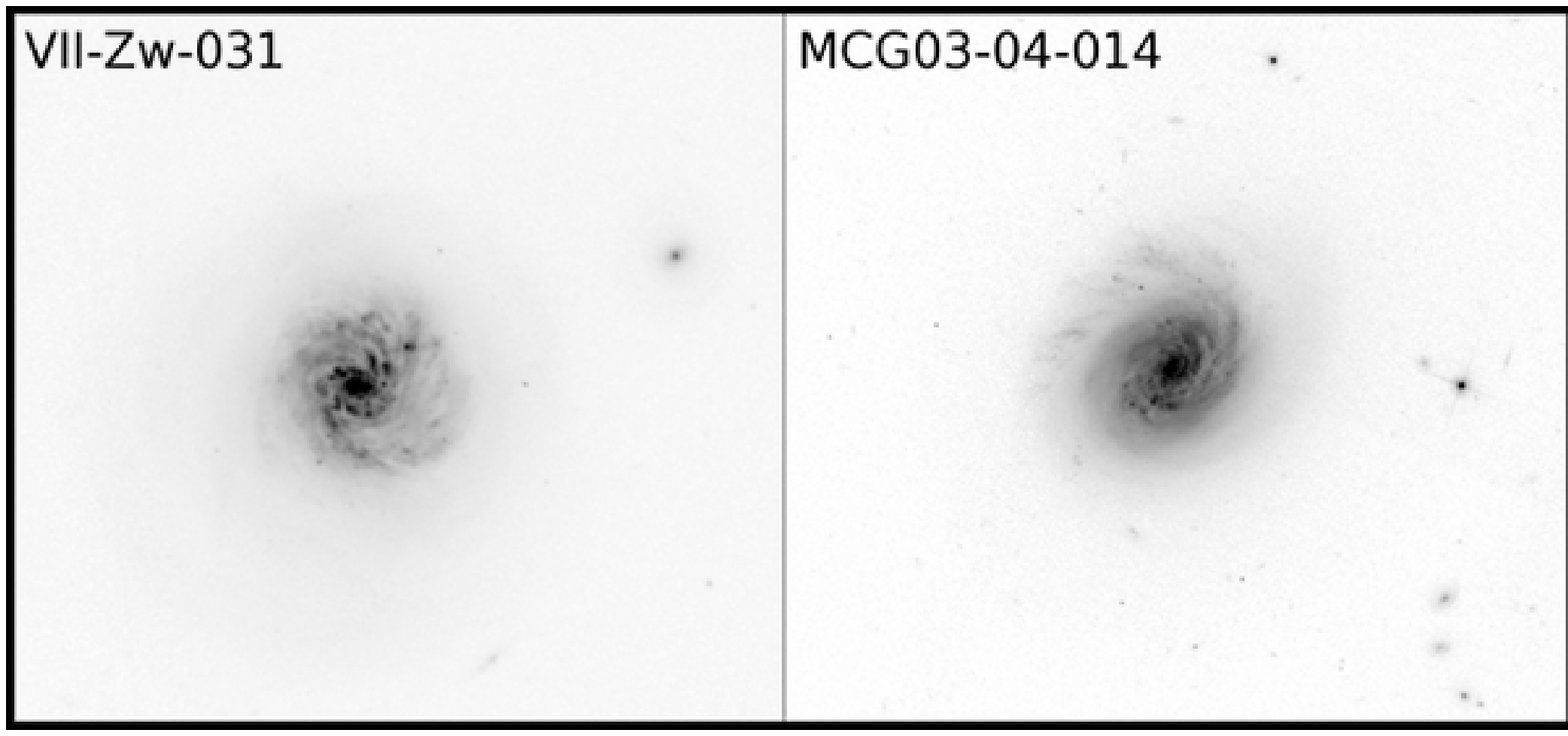}
\end{center}
\caption{The figure at the top shows examples of  I-band images representing the classification in different merger stages, ranging from 1 (top left)  to 6 (bottom right). The stages (see text for details) can be broadly characterized into pre-merging (class 1), merging (2,3, and 4), and post-merging galaxies (5 and 6).  Note that most of our class 1 objects have smaller projected separations than this example (AM07020-601), however, a line-of velocity difference of 63~km~s$^{-1}$ implies that these two galaxies are very likely bound together as merger.
The figure at the bottom shows two examples of undisturbed isolated galaxies. Mergers and isolated galaxies are treated separately thoughout all our studies.}
\label{sequence}
\end{figure}

The non-merger/merger separation as well as the merger classification scheme is based on the wide-field I- and B-band images. While the HST NICMOS H-band images reveal the unobscured nuclei and stellar components, the HST ACS I- and B-band images provide a larger FOV for the detection of companion galaxies and are very sensitive for tracing tidal tails and other interaction features which are essential to identify the merger stage. This provides a distinct advantage over using a simple projected nuclear separation to infer the merger stage, which suffers from degeneracies when exploring a large range in stages. To establish a merger classification we carefully separated at first our sample into galaxies that exhibit interaction features (e.g. tidal tails, bridges) or have nearby companions, and those without any interaction features or nearby companions (undisturbed single galaxies), excluding the latter from our merger sequence analysis. The mergers are classified into six different stages \citep[based on the scheme first proposed by][]{Sur98, Sur98t}; an example for each merger stage based on the I-band images is shown in Fig.~\ref{sequence}: 1 - separate galaxies, but disks symmetric (intact) and no tidal tails, 2 - progenitor galaxies distinguishable with disks asymmetric or amorphous and/or tidal tails, 3 - two nuclei in common envelope, 4 - double nuclei plus tidal tail, 5 - single or obscured nucleus with long prominent tails, 6 - single or obscured nucleus with disturbed central morphology and short faint tails. In sum, these stages may be broadly characterized into pre-merger (1), ongoing merger (2,3,4), and post-merger LIRGs (5,6). 
While for merging stages 2--6 at least two galaxies are obviously interacting or have recently undergone an interaction ( post-merger), this must be not necessarily true for stage 1 as there is no evidence that an interaction (major merger) is responsible for the LIRG activity (see discussion below). However, the maximum line-of-sight velocity difference of  our class 1 mergers is very small (ranging from 65--160~km~s$^{-1}$), which implies that they are very likely bound together as merger.  \par 

\subsection{Double and Multiple Nuclei}
\label{subsec:timescale}

\begin{figure}[ht]
\begin{center}
\includegraphics[scale=0.45]{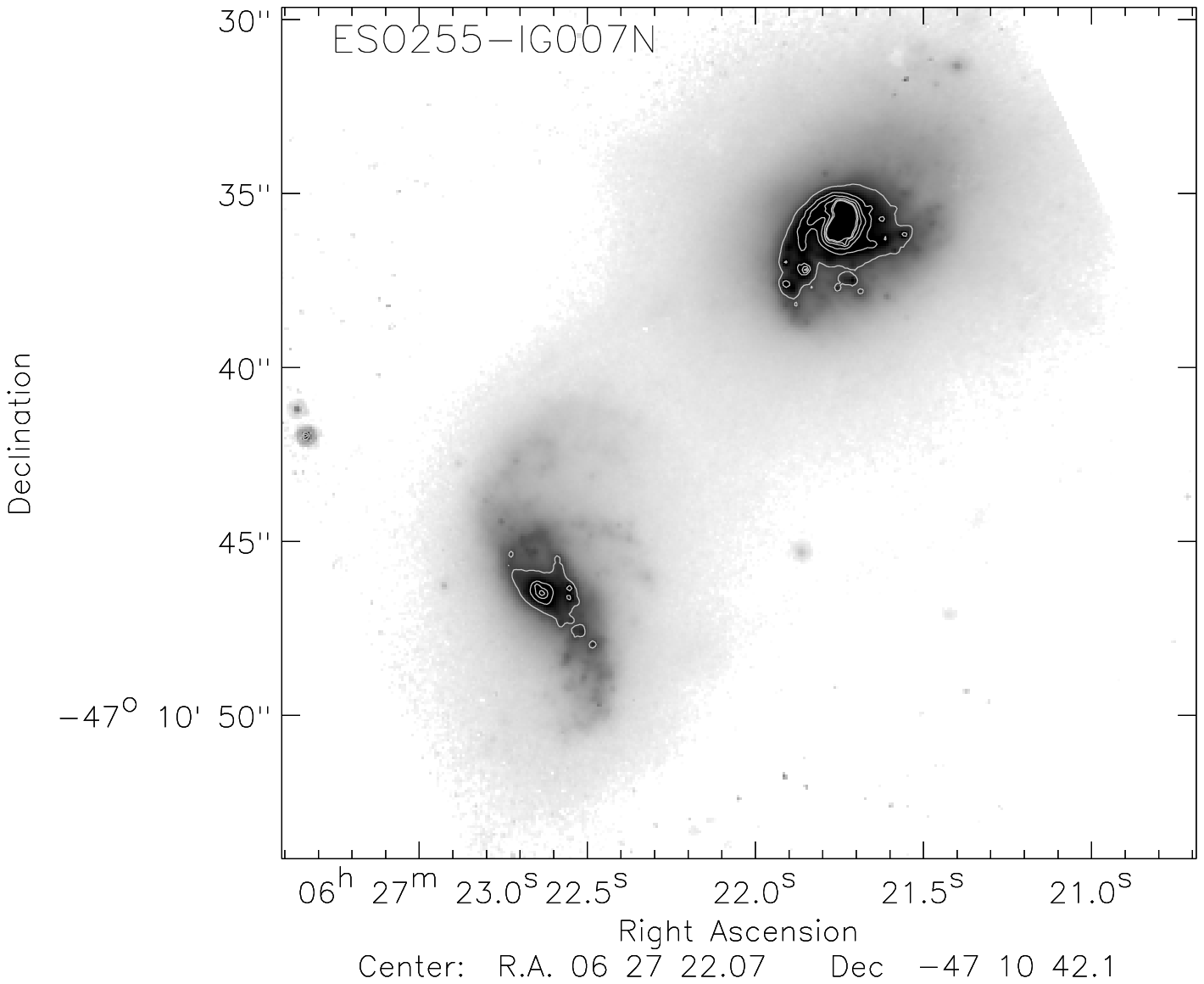}
\includegraphics[scale=0.45]{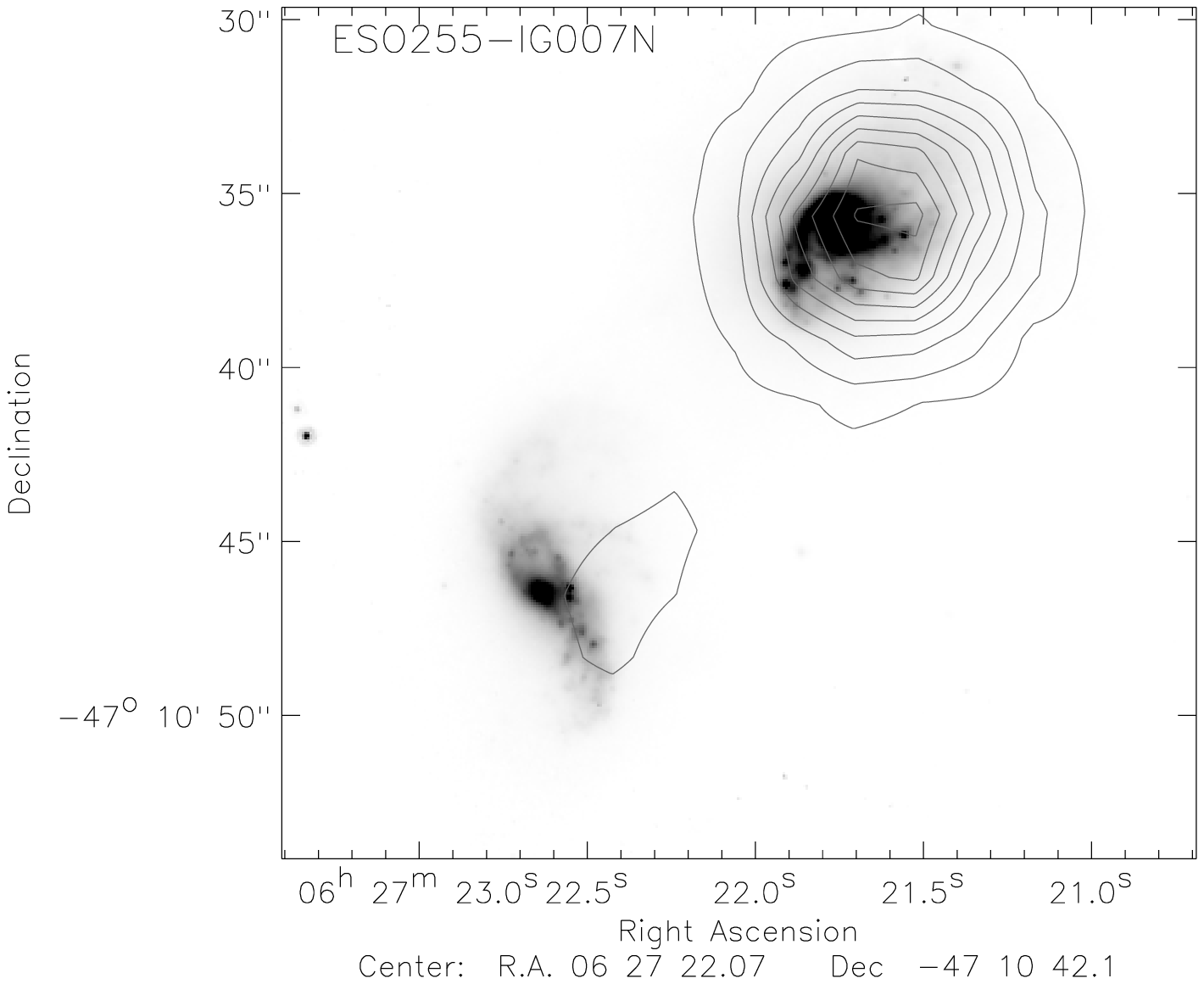}
\includegraphics[scale=0.45]{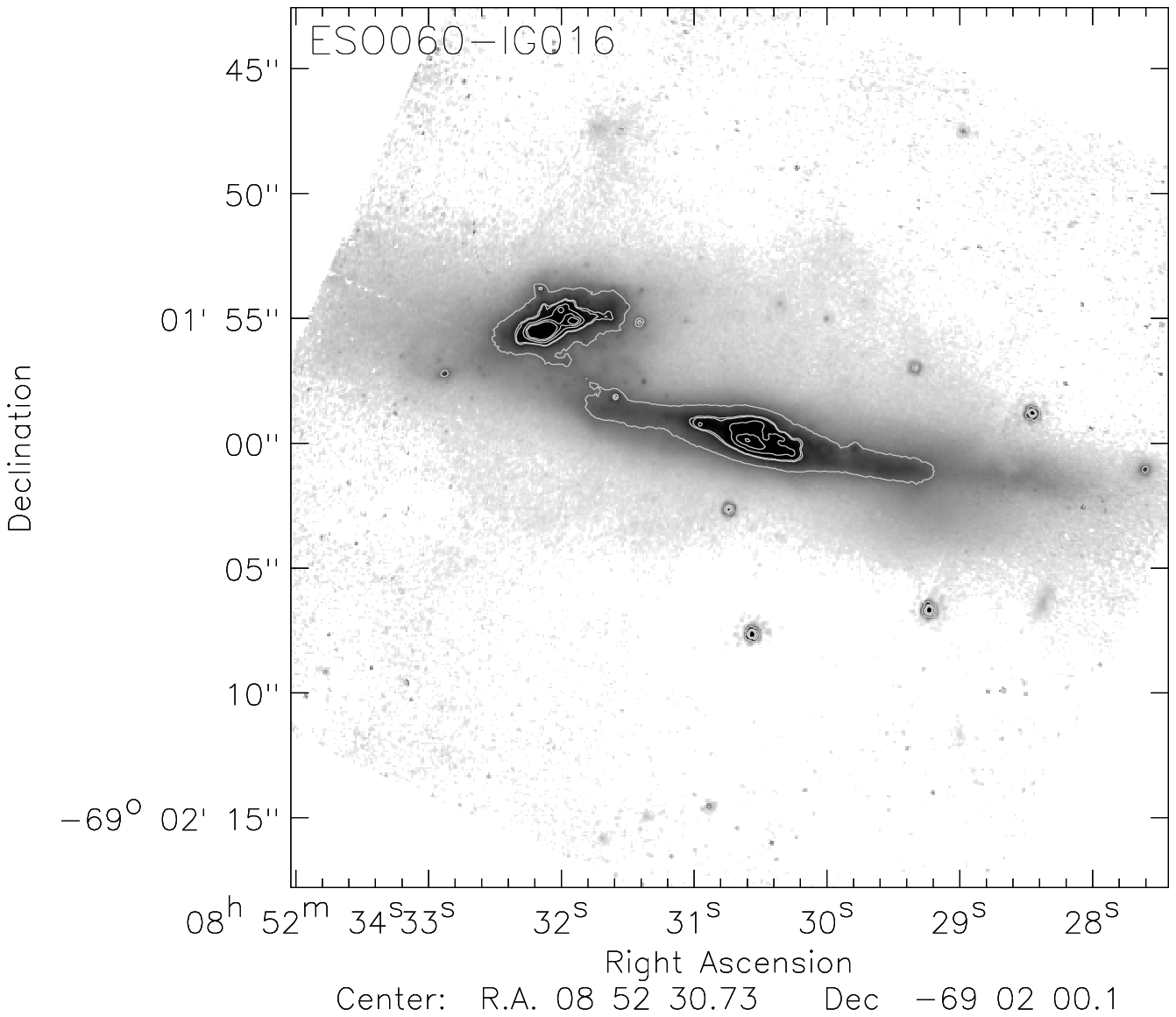}
\includegraphics[scale=0.45]{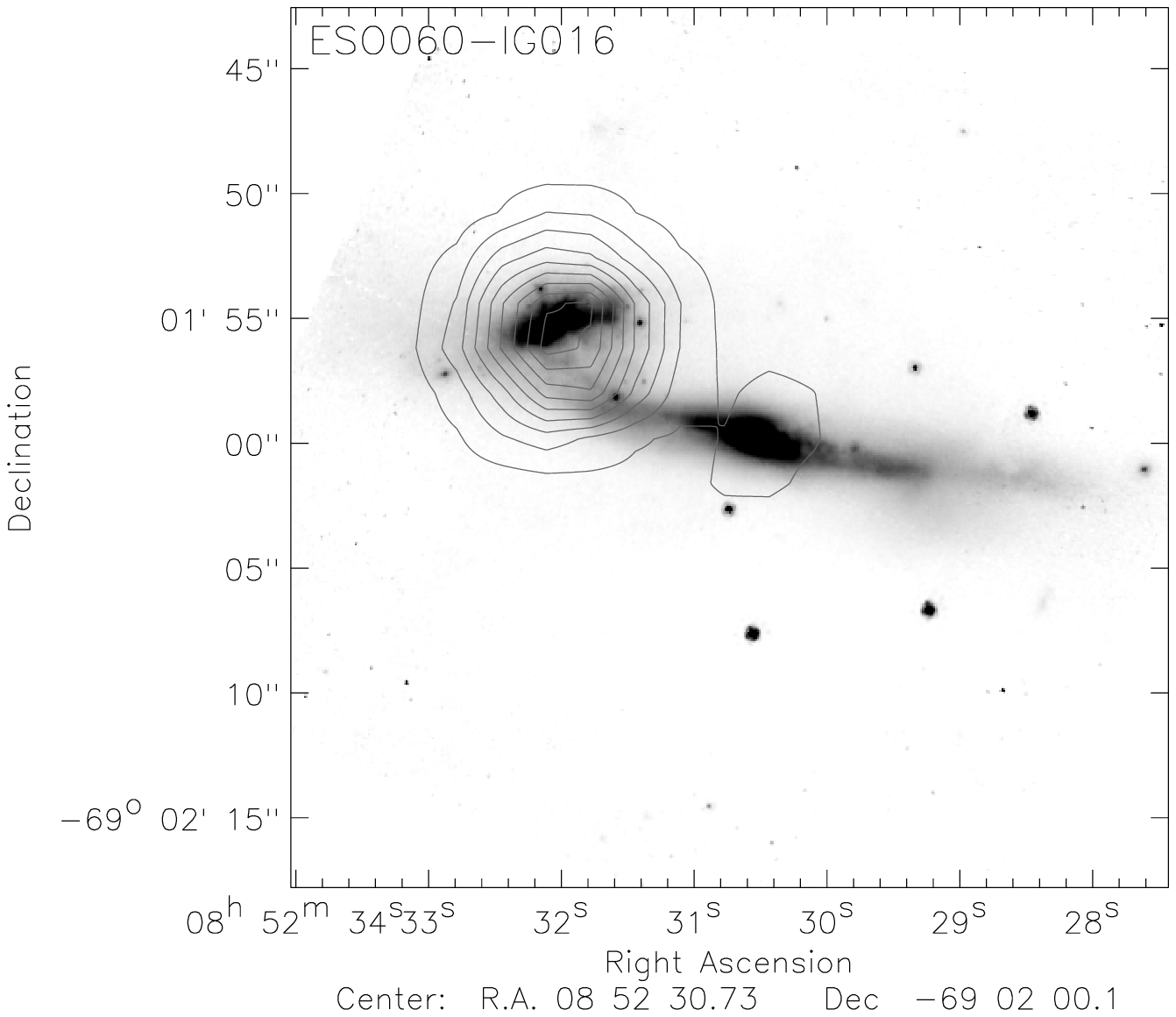}
\includegraphics[scale=0.45]{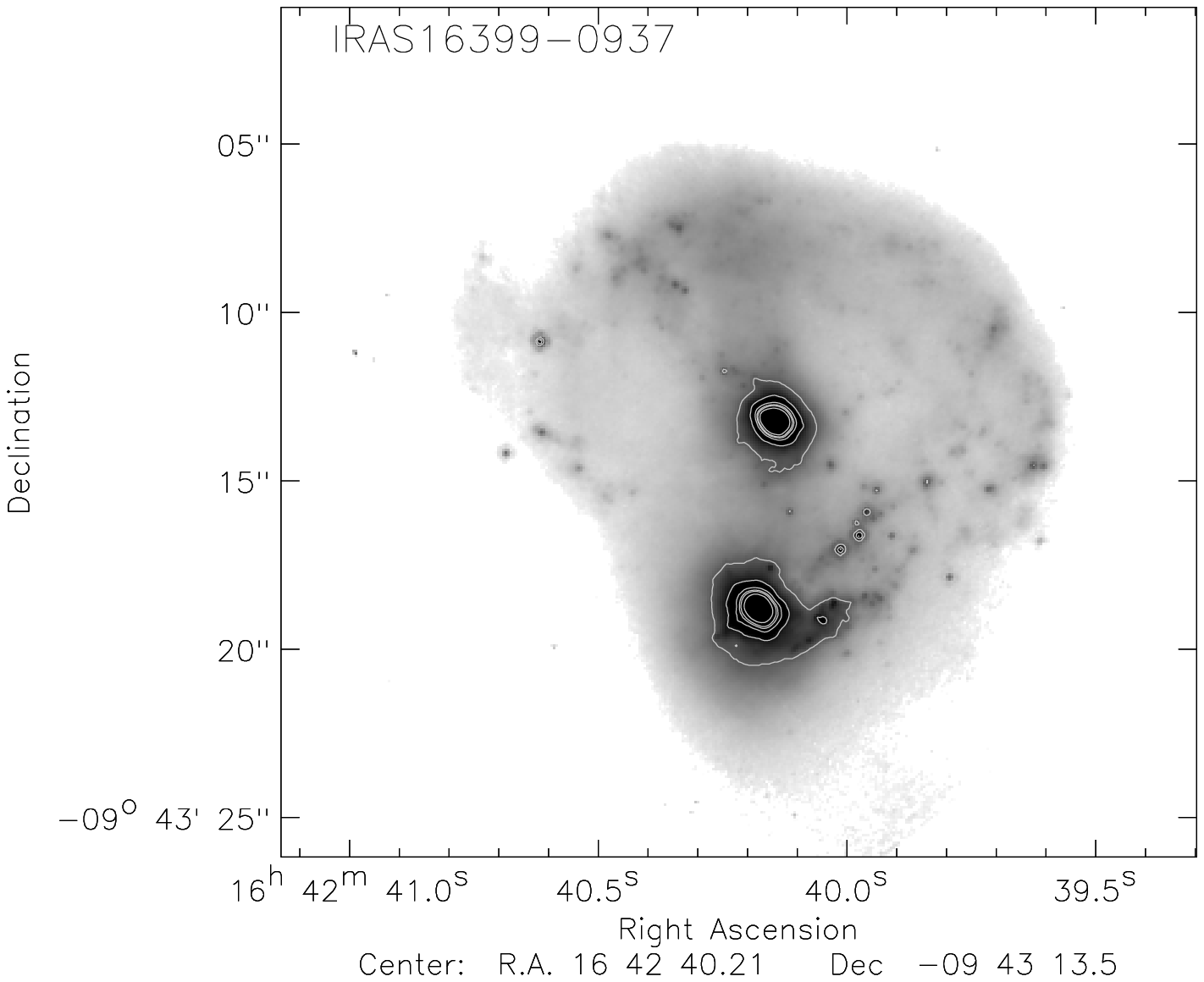}
\includegraphics[scale=0.45]{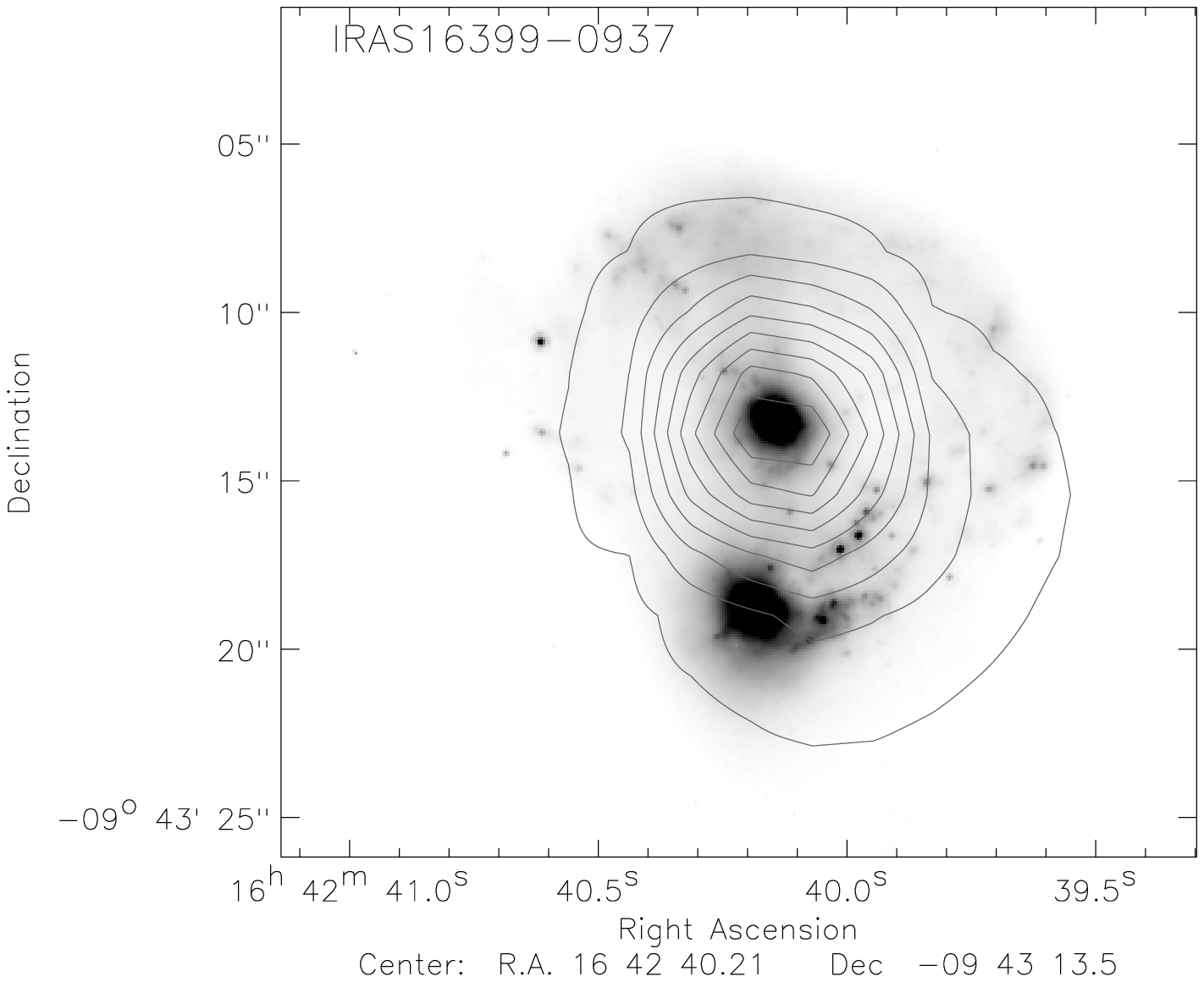}
\end{center}
\caption{Three typical merger examples (top: ESO255-IG007, center: ESO060-IG016, bottom: IRAS~16399-0937) which also show a significant difference in the distribution between the NIR light (left: HST NICMOS F160W images scaled with the square root to reveal also the outer extent of the galaxies and with contours to identify double and multiple nuclei) and the mid-IR light (right: HST NICMOS image scaled linear and overlaid with MIPS 24~$\mu$m contours). Figures for the rest of the sample are available in the online version of the Journal.}
\label{NICMOS_MIPS}
\end{figure}

We find double as well as multiple nuclei with a large range of nuclear separations (100~pc to $>$10 kpc). Fig.~\ref{NICMOS_MIPS} shows three typical merger examples of our NICMOS images including a comparison with the mid-IR emission as observed with 24$\mu$m MIPS images (Figures for all LIRGs are available in the online version of the Journal). We find that a significant fraction of our sample shows sign for late merger stage which is typically characterized by tidal tails (see \S.~\ref{subsubsection:sequence} for more details), and in some cases, by a disturbed center (with eventually unresolved double nuclei). \par

\clearpage       

To understand how star clusters, AGN, and nuclei evolve as a function of luminosity and merger stage, it is essential to first accurately count and measure the central structures. Although NIR images trace the old stellar nuclear structure, the possibility of NIR emission from accompanying intense star-forming regions rather from a secondary nuclei cannot be ruled out (see \S.~\ref{subsec:dis-AGN} for more detailed discussion). We carefully checked different contour levels for each galaxy to identify the nuclear structure and possible associated galaxy disks. Figures with HST NICMOS contour levels for all LIRGs are available in the online version of the Journal. To estimate the number of systems with at least two nuclei we followed a conservative approach. We count all LIRGs in our sample that have primarily two galaxies with each of them exhibiting a (eventually disturbed or merging) disk as well as LIRGs where no clearly visible disk could be associated with every nucleus, but exhibiting two significant separated NIR light concentrations (with spherical/elliptical shape expected from nuclei) with NIR luminosity ratios $\lesssim 3:1$.\par 

To estimate the number of galaxy pairs that are not within the NICMOS Field of View (FOV), we take into account galaxy pairs that are visible in the optical HST ACS images (F814W filter, I  band) that show signs for interaction such as tidal tails. Out of the 73 systems, five were interacting systems with one of the companions outside the NICMOS FOV. We also find 5 LIRGs with galaxy pairs within 100~kpc that show no strong interaction features, and hence are not considered as interacting LIRG systems in our analysis. 
Note that up to 10\% of galaxies in our sample that are not classified as double nuclei systems show a (slightly) disturbed center (likely caused by a minor merger or strong secular evolution effects such as bars or nuclear spirals) but not two separable NIR light concentrations that are typical for double nuclei.  \par

\begin{figure}[ht]
\begin{center}
\includegraphics[scale=0.7]{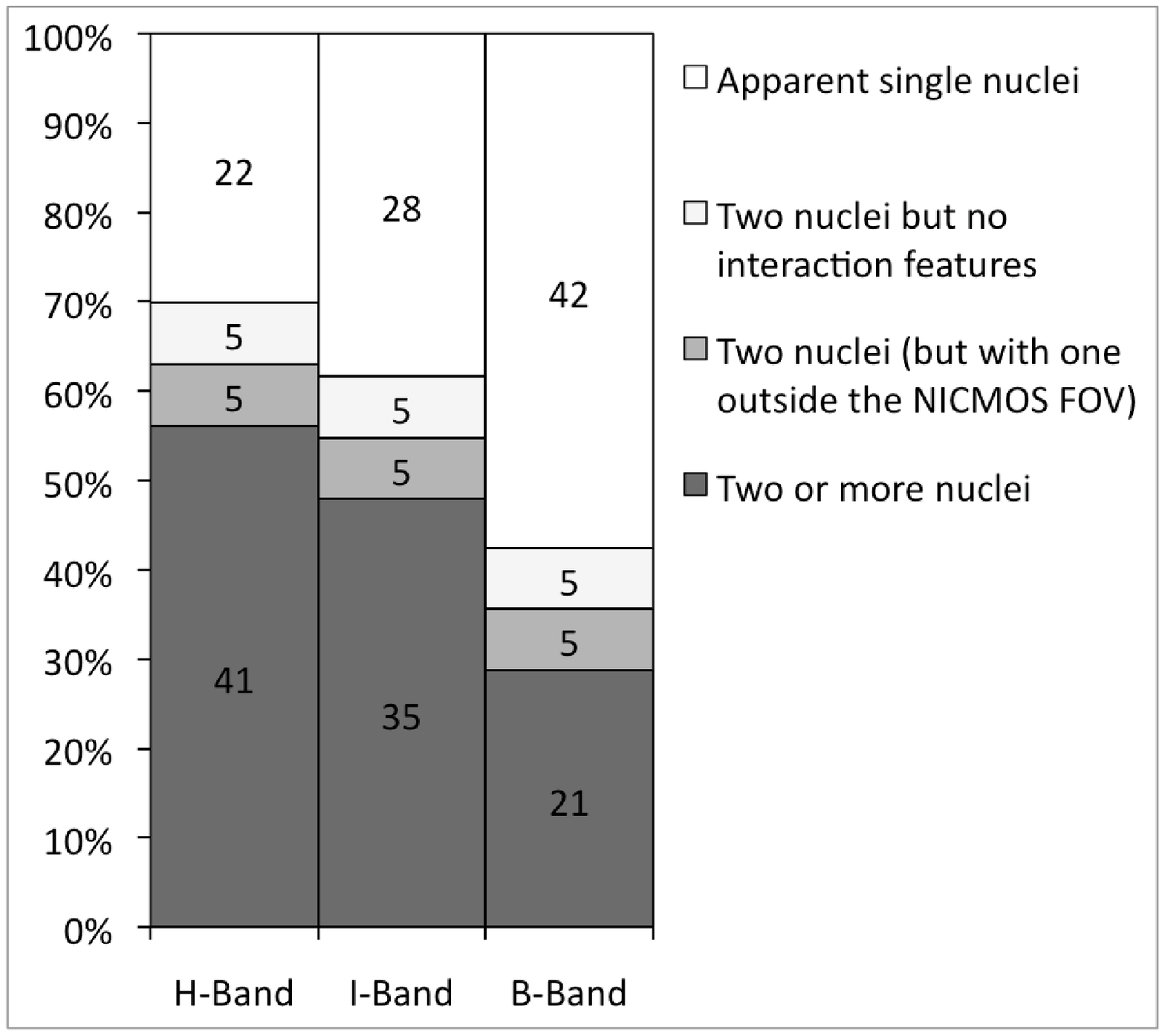}
\end{center}
\caption{Comparison of the number of LIRG with multiple nuclei detected in the H, I and B bands. From bottom to top: (1) The number of systems with at least two nuclei in galaxies that show interaction features (in dark grey), (2) with two galaxies (visible in I and B band) but only one of them within the NICMOS FOV (in grey; we assume that galaxies visible in the I band would also been seen in the H band), (3) with double nuclei whose host galaxies show no sign for interaction (neither in the H, I, or B band images; in light grey), and (4) galaxies with apparent single nuclei ($\Delta R\lesssim 150$pc, in white) and major merger interaction features . These results show that half of the double nuclei are obscured by dust as they are not visible in the B band. Triple nuclei are found in at least 3 (possibly 5) LIRG systems ($\sim$6\%).}
\label{nuclei_H-B}
\end{figure}

\begin{figure}[ht]
\begin{center}
\includegraphics[scale=0.6]{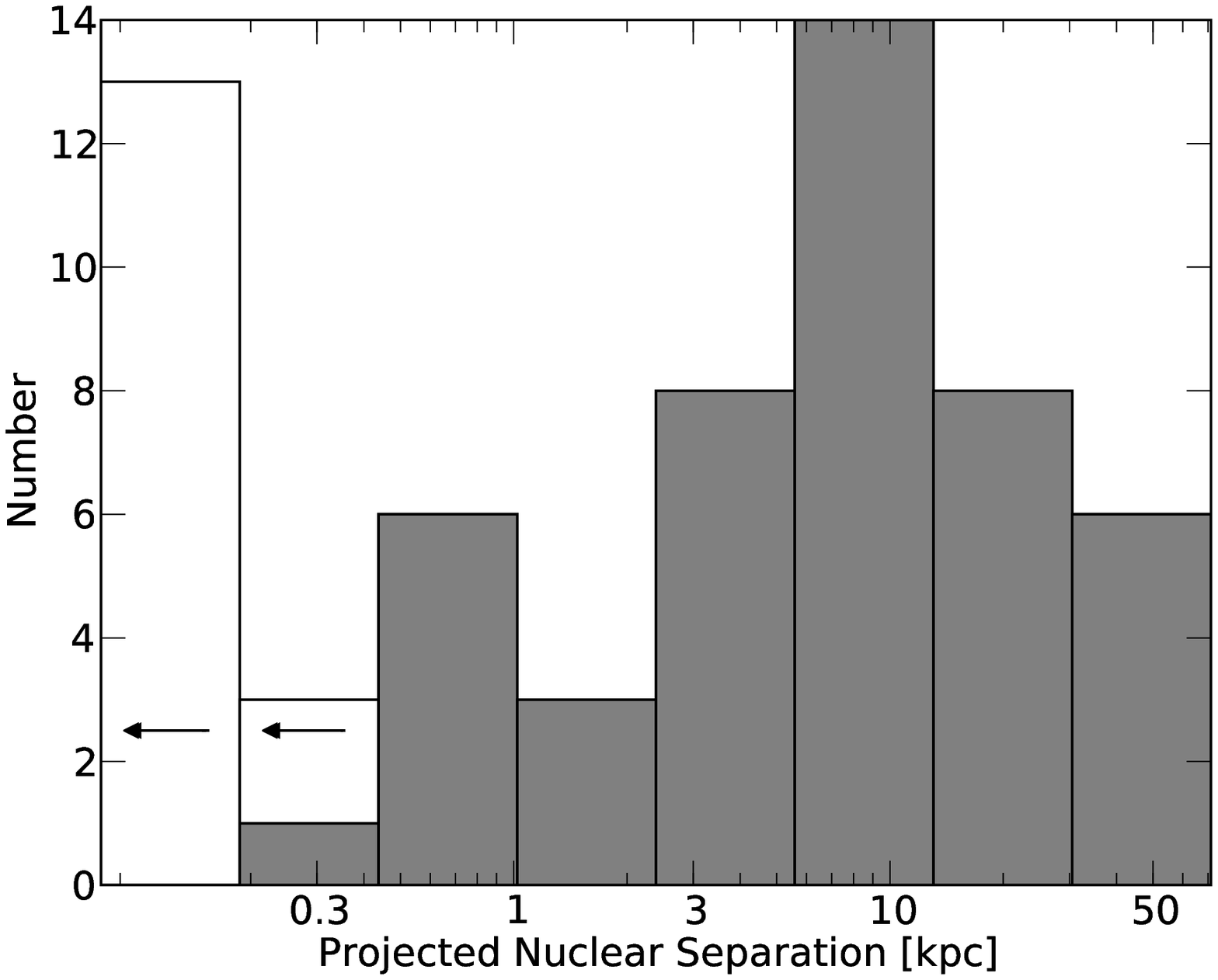}
\end{center}
\caption{Histogram of the projected nuclear separation between double nuclei. Fourteen apparent single nuclei that exhibit prominent interaction features, such as tidal tails, are also included in white, converting our angular resolution limit of  $0.15\arcsec$ to the linear size at their corresponding distance. For triple nuclei systems, the smallest projected nuclear separation is chosen.}
\label{hist_dist}
\end{figure}

We find that the fraction of LIRG systems (log[L$_{IR}$/L$_\odot$] $=$ 11.4 $-$ 12.5 ) with at least two interacting nuclei is 63\% (see Fig.~\ref{nuclei_H-B}), with a median projected separation (see Fig.~\ref{hist_dist}) of 7.2~kpc.  If we also take into account apparent single nuclei in galaxies that exhibit tidal tails (given the spatial resolution of $0.15\arcsec$ as upper limit of the projected nuclear separation), the median projected separation is 4.9~kpc. Six double nuclei that are visible in the H-band can not be identified as two separated nuclei in comparable resolution HST I-band images (ACS F814W filter), likely due to dust obscuration (not visible B-band images as well). 
Triple nuclei are found for ESO~255-IG007, II~ZW~096, NGC~6670, and possibly UGC~02369 and NGC~3690. We note that the fraction of triple nuclei systems in our sample ($\sim$6\%) is similar to the percentage of triple systems ($\sim$5\%) found in a population of $\sim$11,000 galaxies in the Local Supercluster and vicinity \citep{Mak09}. 
\par

\begin{figure}[ht]
\begin{center}
\includegraphics[scale=1.0]{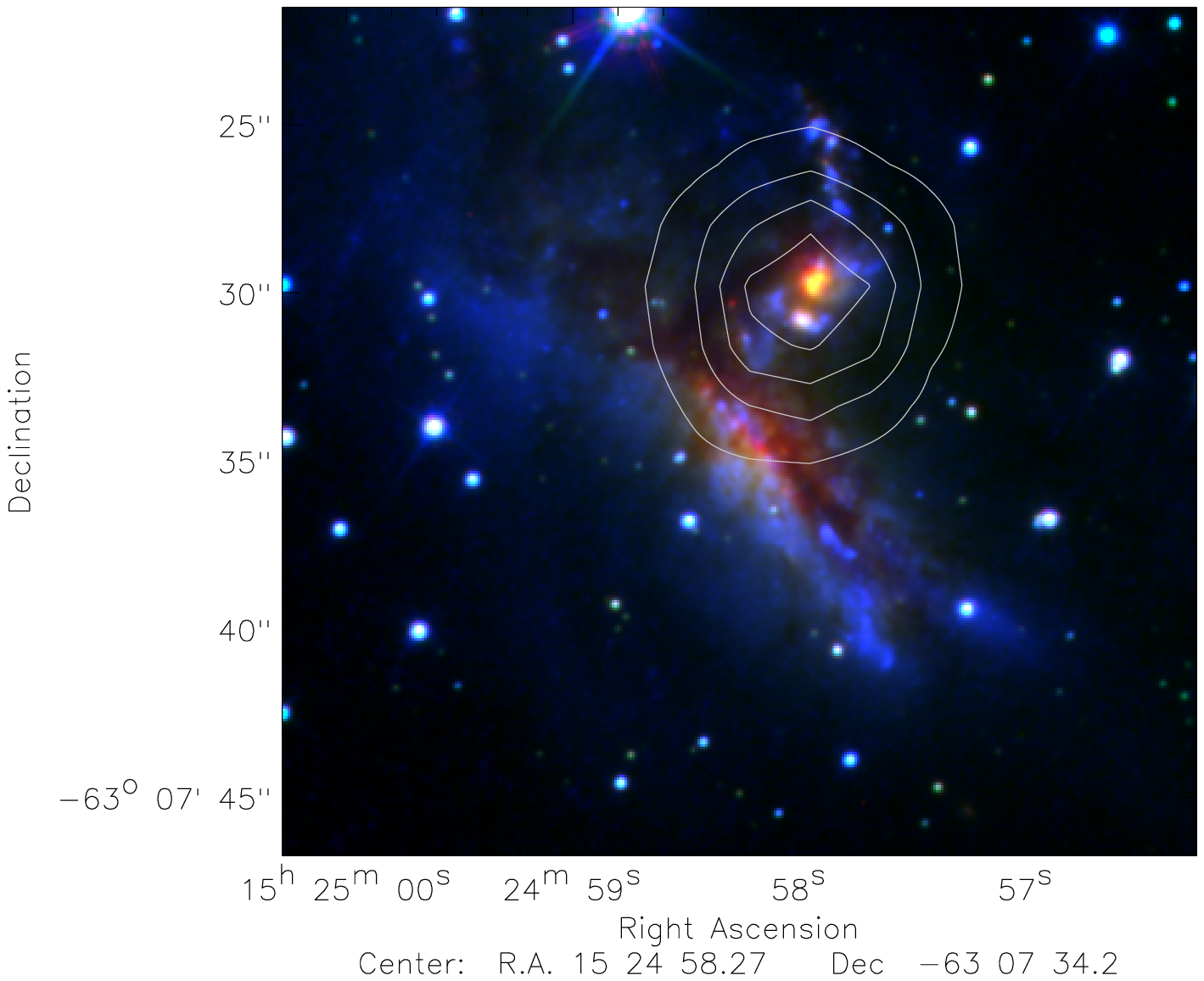}
\end{center}
\caption{A three-color composite image (red: H-band, green: I-band, blue: B-band) with mid-IR (24~$\mu$m) light emission (contours) for ESO~099-G004. This example clearly shows the importance of NIR images to reveal multiple nuclei that are not visible in B-band images (blue) due to dust extinction although dominating the mid-IR light emission (contours).}
\label{eso099}
\end{figure}

\begin{figure}[ht]
\begin{center}
\includegraphics[scale=1.0]{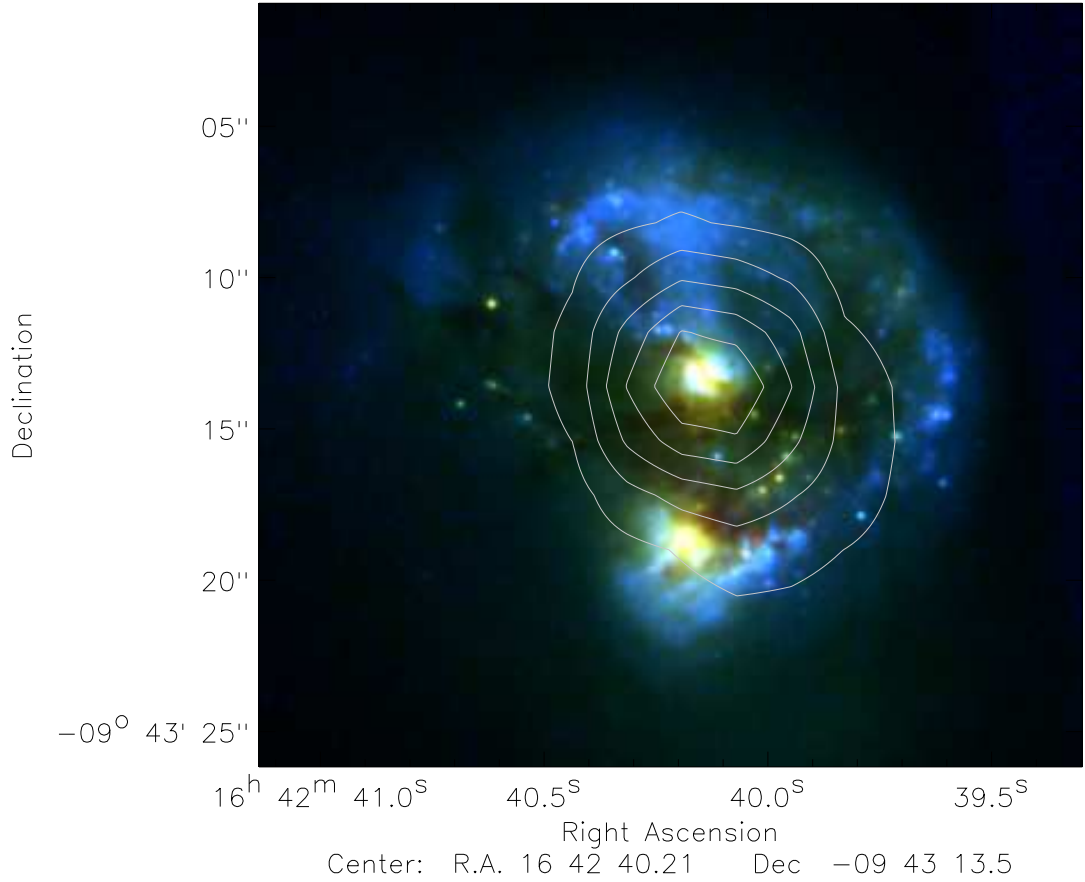}
\end{center}
\caption{IRAS~16399-0937; three-color composite image (red: H-band, green: I-band, blue: B-band) with MIPS 24~$\mu$m contours. This example shows two nuclei (separation distance $\sim$3~kpc) of roughly the same size and luminosity in the H-band, albeit most of the mid-IR emission ($>$90\%) originates only from one of the nuclei.}
\label{iras16339}
\end{figure}

To obtain the number of nuclei that are obscured by dust, we compare the NICMOS images with comparable resolution HST B-band images (ACS F435W filter) and find that only $\sim$50\% of the double nuclei are visible in the B-band (see Fig.~\ref{nuclei_H-B}). In detail, the fraction of LIRG systems with at least two nuclei visible in the B-band within the NICMOS FOV is 28\%. Excluding the NICMOS FOV limit we obtain a fraction of 35\% LIRG systems with interacting double nuclei visible in the B-band. Interestingly, almost half of the dust-obscured nuclei (10 out of the 20 that are seen in H-band but not in B-band) are responsible for the bulk of the mid-IR emission (MIPS-24$\mu$m).
A typical example is ESO~099-G004 which shows two nuclei in the H-band with a projected nuclear separation of $\sim$3.6~kpc, neither of which can be seen in the B-band (see Fig.~\ref{eso099}). Interestingly, all of the mid-IR emission is emitted from one nucleus in this double nuclei system. Another example is IRAS~16339-0937, which shows clearly two nuclei (separation distance $\sim$3~kpc) of roughly the same size and luminosity in the H-band, albeit most of the mid-IR emission ($>$90\%) originates only from one of the nuclei. The true nature of this source is not evident in the B band data (see Fig.~\ref{iras16339}).\par

\clearpage

\subsection{GALFIT Results}
\label{subsec:res-GALFIT}

\subsubsection{GALFIT Analysis}
\label{sec:galfit}

To derive detailed information about the structural properties of LIRGS we performed a 2-dimensional decomposition of all galaxies in our NICMOS sample using GALFIT \citep{Pen02,Wei09}. As most of the galaxies in our sample are mergers, and hence exhibit disturbed outer disks, we have not attempted to obtain a precise model for the entire merger system as arbitrary fitting of the high spatial frequency components does not add to a parametric model of the systems. Instead, we focused on an accurate measurement of the bulge luminosity and central concentration indices, as these parameters provide important information about the build-up of stellar mass in the center of galaxies during the merger process.  We performed a careful analysis of each galaxy, also identifying cases where a global bulge-disk decomposition would not work. \par

Running GALFIT on an image requires initial preparation. The desired fitting region and sky background must be provided, and the PSF image, bad pixel mask (if needed), and pixel noise map must be generated. Our analysis steps are described as follows: (1) we converted the integrated galaxy counts per second to an apparent Vega magnitude using the NICMOS data calibration description (Dickinson et al. 2002):
\begin{equation}
m = -2.5 ~ log(F_{CPS} \times PHOTFNU/F_{\nu}(Vega)),
\end{equation}
with the source flux $F_{CPS}$ in counts per second, the Vega flux $F_{\nu}[VEGA]$) (for NIC2 at F160W, $F_{\nu}[Vega]$
is 1043.6 Jy), and PHOTFNU as conversion factor from counts per second to Jansky ($1.49585 \times 10^{-6}$ Jy s count$^{-1}$). (2) The fitting region must be large enough to include the outer galaxy disk or entire merger system, as well as some sky region. (3) Sky backgrounds were measured separately using a region where no emission is present and designated as fixed parameters for all GALFIT procedures. In general a fitting of the sky background is not recommended as the wings of the bulge S\'{e}rsic profile \citep{Ser68} can become inappropriately extended, resulting in a S\'{e}rsic index that is too high. (4) We simulated the required PSF image with the TinyTim code \citep{Kri97} using a pixel scale identical to that of the observations.  (5) The uncertainty images are generated by the HST reduction pipeline and used as weighting images required for GALFIT to perform the $\chi ^2$ minimization.\par

After these preparation steps we carried out an iterative process to perform a 1-3 component decomposition of the NICMOS images described in the following: GALFIT requires initial guesses for each component it fits and uses a Levenberg-Marquardt downhillgradient algorithm to determine the minimum  $\chi ^2$ based on the input guesses. At first we fitted single S\'{e}rsic profiles that are centered on the galaxies nuclei (typically 1-2 nuclei per image). For merger systems that consists of two or more galaxies, we take advantage of GALFITs ability to fit simultaneously multiple components at different spatial regions.
The S\'{e}rsic profile has the following form:
\begin{equation}
\Sigma = \Sigma_e e^{-k[(R/R_e)^{1/n_s}-1]}
\label{eq:sersic}
\end{equation}
with the effective radius $R_e$, the surface brightness $\Sigma_e$ at $R_e$, the power law index $n_s$, and $k$ coupled to $n_s$ such that half of the total flux is always within $R_e$.
The advantage of the S\'{e}rsic profile is that it is very useful for modeling different components of a galaxy such as bulges, flat disks, and bars, simply by varying the exponent. As the index $n_s$ becomes smaller, the the core flattens faster for $R < R_e$, and the intensity drops more steeply for $R > R_e$. 
Classical bulges typically have a S\'{e}rsic index of $n=4$ (in general known as de Vaucouleurs profile), but can vary between $1<n<10$, while the outer disk follows an exponential profile ($n=1$).\par 

The GALFIT output parameters based on a single S\'{e}rsic profile are used as initial parameters for the multi-component fitting and are applied as a residual image in the cases that more than one galaxy is present per image. The center coordinates have been fixed for the subsequent fitting steps. A multi component fit is performed on each individual galaxy. In general a double S\'{e}rsic component is used, while possible additional galaxies are subtracted using their single S\'{e}rsic parameters from step 1. In cases where the galaxy shows no significant disturbances we fitted the sum of an exponential and S\'{e}rsic component with fixed position angle. An additional S\'{e}rsic component has been fitted when a bar was present. As suggested by \cite{Pen02}, the bar can be well described by an elongated, low index S\'{e}rsic (n $<$ 1) profile. For each galaxy a variety of different multi-component fits are performed (typically 4 - 7 different fit setups per galaxy) and compared afterwards. 
Fitted models were selected only as long as the model parameters were all well behaved, i.e. requiring that all parameters such as bulge radius and S\'{e}rsic index converged within the range of reasonable values (e.g. $n_s<10$). The best fit was chosen based on several criteria: First, GALFIT calculates the $\chi^2$ for each model which generally declines as more free parameters or S\'{e}rsic components  are given. However, a small $\chi^2$ does not necessarily mean that the solution with an extra bar component is more correct physically. Thus, an increasing $\chi^2$ for a model with an extra bar component was interpreted that the fit should not be adopted, but a decreasing $\chi^2$ was not considered as a sufficient condition to adopt an extra bar component. In cases with prominent bars, a symmetric light distribution due to unsubtracted bar light was often found in the bulge-disk residuals. For these cases we added an extra bar component since it would otherwise lead to an extended bulge.\par 

For objects with central point sources such as an AGN, the bulge S\'{e}rsic index can be overestimated unless an extra nuclear component (PSF component) is added to the model. On the other hand, nuclear stellar cusps can mimic a central point source and adding a nuclear PSF would lead to an underestimation of otherwise a steeper S\'{e}rsic index. Therefore, we used GALFIT in a first attempt without a central PSF component and calculated the central residual excess light from the residual image of the final fit. In cases where a central residual was present (for $\sim$10\% of the sample) we reran GALFIT with an additional PSF component to take into account the extra central light originating from a central point source. We also checked all LIRGs that show signs for AGN activity (based on 6.2~$\mu$m PAH Equivalent Width, see \S~\ref{subsec:dis-AGN}) whether a PSF component is present or not. For those cases where the model successfully converged with the extra component, the images were visually inspected to verify the presence of a central bright source. Where new model parameters were not unreasonable and not identical to the case without the point source, the new model was adopted. This was the case for 14 galaxies (17\% of the sample), but only for 9 of them reasonable bulge parameters (with $L_\mathrm{Bulge}$, $R_\mathrm{Bulge}$, and $n_s > 3\sigma$) could be obtained. For consistency we also checked galaxies with less significant residual light and larger PAH EQWs (no significant AGN contribution), but an extra nuclear component was not found for any of those galaxies.

\subsubsection{Derived Bulge Parameters and Black Hole Mass Estimation}

\begin{figure}
\begin{center}
\includegraphics[scale=0.53]{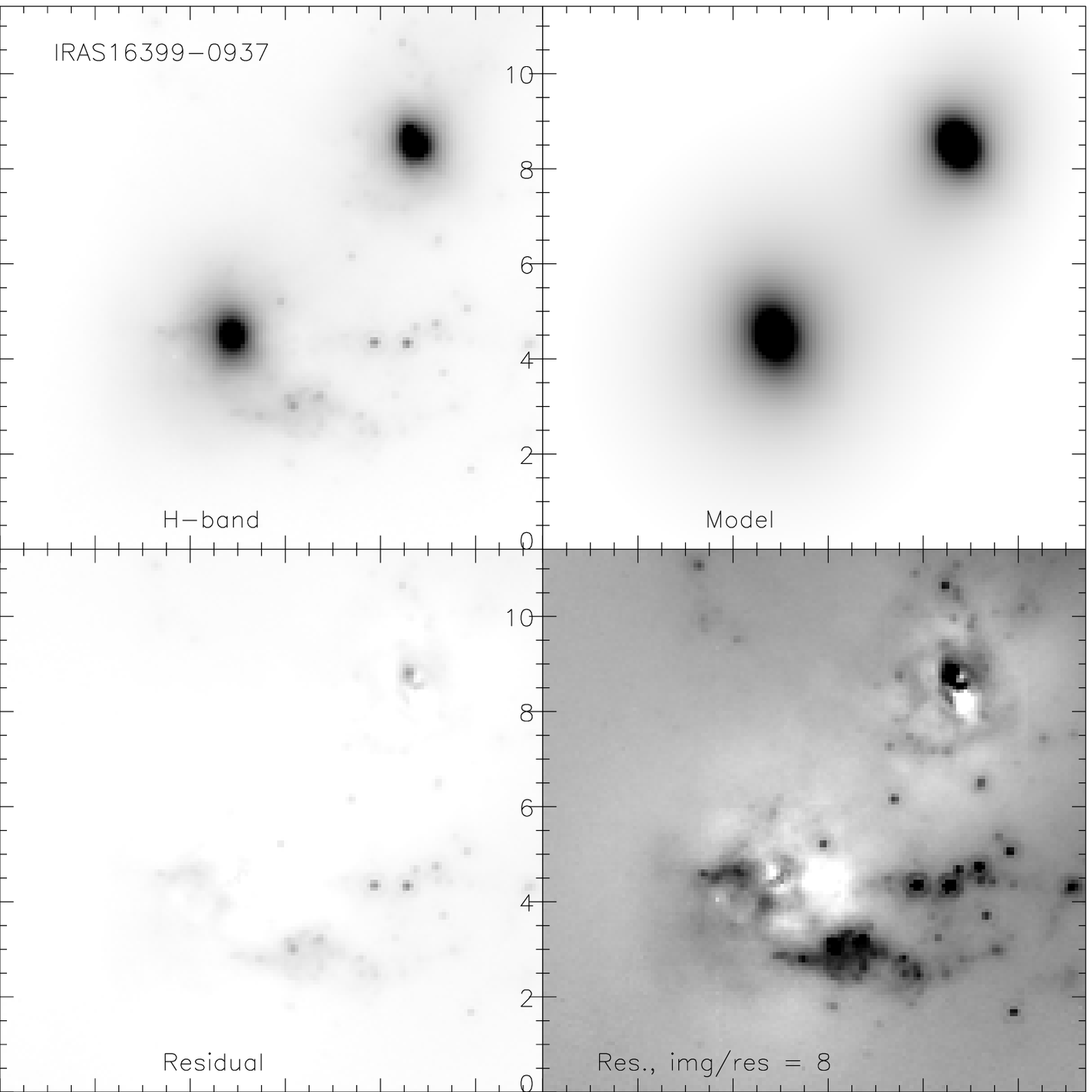}\\
\bigskip
\includegraphics[scale=0.53]{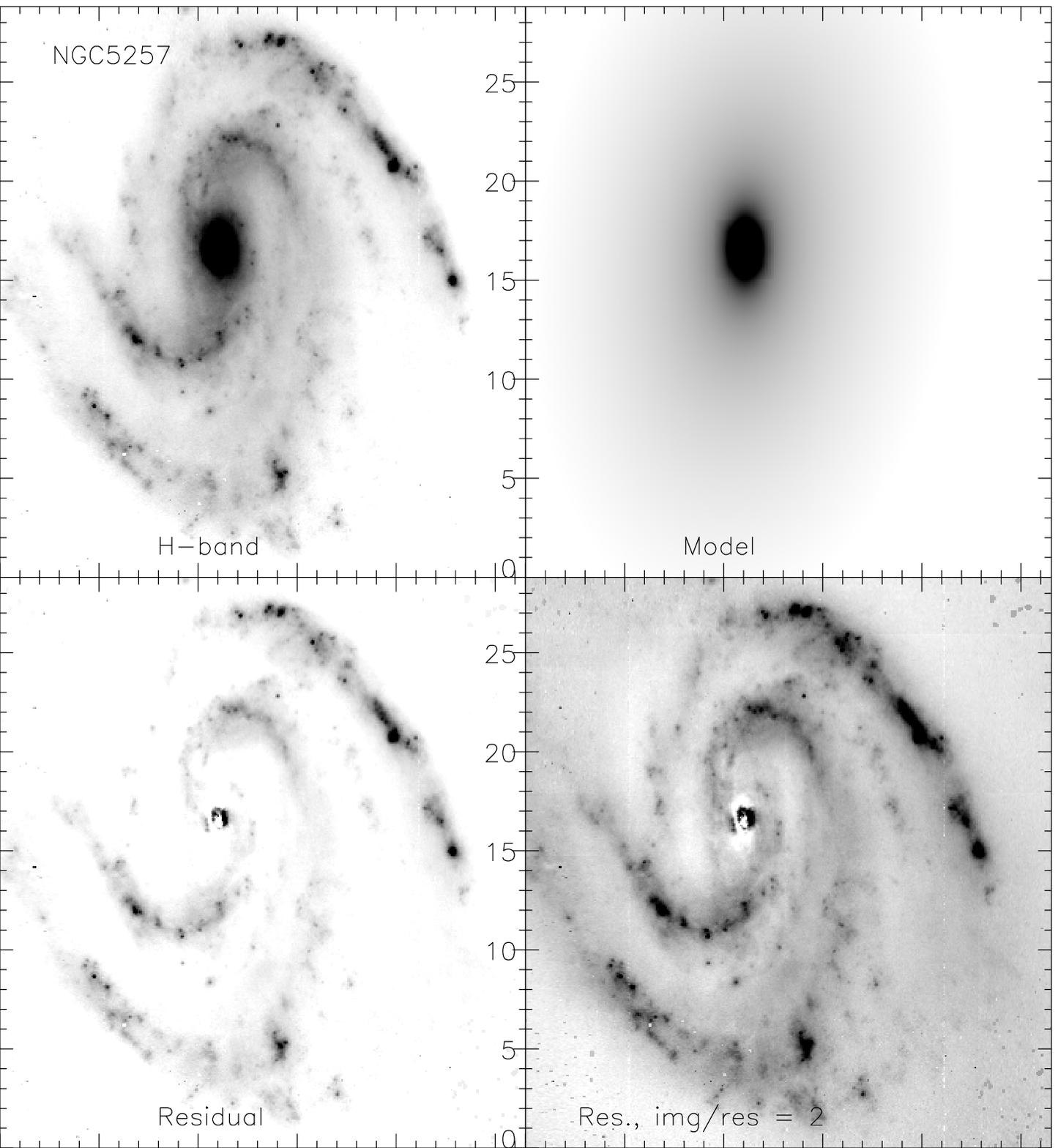}
\caption{GALFIT output images. Top left: Region of NICMOS image that is fitted (box correspond to fitted region), Top right: Galfit model, Bottom left: Residual image (shows the difference between model and NICMOS image) with the same brightness level as the NICMOS image, Bottom right: Residual image scaled to its maximum brightness (the brightness ratio of NICMOS image to scaled residual map is shown as well at the bottom) to highlight the faint diffuse emission, spiral structure, and clusters remaining in the model-subtracted data. The axes are in scales of arcsec and the figures are orientated in the observational frame. Figures for the rest of the sample are available in the online version of the Journal.}
\label{galfit}
\end{center}
\end{figure}

The following morphological parameters are derived via fitting of our H-band images using GALFIT (see \S~\ref{sec:galfit}): (1) the total bulge luminosity $L_\mathrm{Bulge}$, (2) the S\'{e}rsic index $n_s$ of the bulge component radial profile, and (3) the bulge radius $R_\mathrm{Bulge}$ (see Tab.~\ref{tab:galfit}). We have obtained bulge parameters for 73 nuclei (with $L_\mathrm{Bulge}$, $R_\mathrm{Bulge}$, and $n_s > 3\sigma$) in our sample and calculated the central black hole masses using the Marconi-Hunt relation \citep{Mar03} given as
\begin{equation}
\mathrm{log} M_{BH} = 8.19(\pm 0.07) + 1.16(\pm 0.12) \cdot [\mathrm{log}(L_{H,Bulge}) - 10.8]
\end{equation}   
with the bulge luminosity $log(L_{H,Bulge})$ measured from the H-band. Moreover, we calculated the bulge luminosity surface density defined as $L_\mathrm{Bulge}/R_\mathrm{Bulge}^2$, and the ratio of the core excess luminosity to bulge luminosity  $L_\mathrm{excess}/L_\mathrm{Bulge}$ (measured within the central $\sim30-300$~pc ). The central residual light or core excess luminosity $L_\mathrm{excess}$ is derived from the center of the residual image (in detail the bulge model subtracted H-band image). In case of a significant central residual light we subsequently refined the core excess luminosities by fitting a central PSF component in GALFIT (see \S.~\ref{sec:galfit}). Two simple examples of GALFIT models and residual images are shown in Fig.~\ref{galfit} (GALFIT model and residual images for all LIRGs are available in the online version of the Journal).

\begin{figure}[h]
\begin{center}
\includegraphics[scale=0.52]{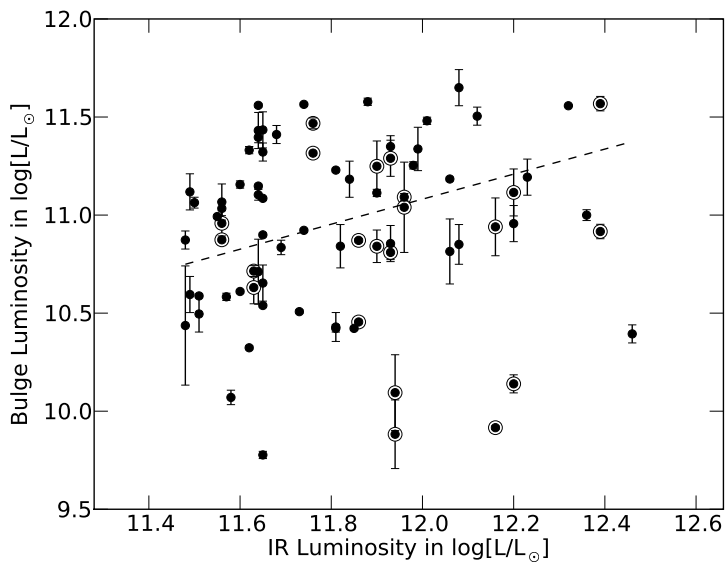}
\end{center}
\caption{Bulge luminosity derived from GALFIT versus total IR luminosity \citep[using the flux densities reported in the RBGS,][]{San03}. Bulges that belong to the same merger system are highlighted with an outer ring at a common IR luminosity. A linear fit over all data points (dashed line) reveals $\mathrm{log}(L_{Bulge}/L_{\odot}) =3.84(\pm2.14) + 0.61(\pm0.18)\cdot \mathrm{log}(L_{IR}/L_{\odot})$.}
\label{bulge_IRlum}
\end{figure}

We find only a small trend ($\sim 3 \sigma$ significance level, see Fig.~\ref{bulge_IRlum}) in which the bulge luminosity is correlated with the total IR luminosity \citep[using the flux densities reported in the RBGS,][]{San03} as given by the following fit parameters:  
\begin{equation}
\mathrm{log}(L_{Bulge}/L_{\odot}) =3.84(\pm 2.14) + 0.61(\pm0.18)\cdot \mathrm{log}(L_{IR}/L_{\odot})
\label{eq:IR_bulge}
\end{equation}
This result indicates a small dependence between bulge stellar mass and total IR luminosity since a possible contribution of AGN activity on the bulge NIR luminosity is unlikely (as discussed in \S~\ref{subsec:dis-AGN}). 

\subsubsection{Bulge Properties as a Function of  Merger Stage}
\label{subsubsection:sequence}

\begin{figure}[h]
\begin{center}
\includegraphics[scale=0.7]{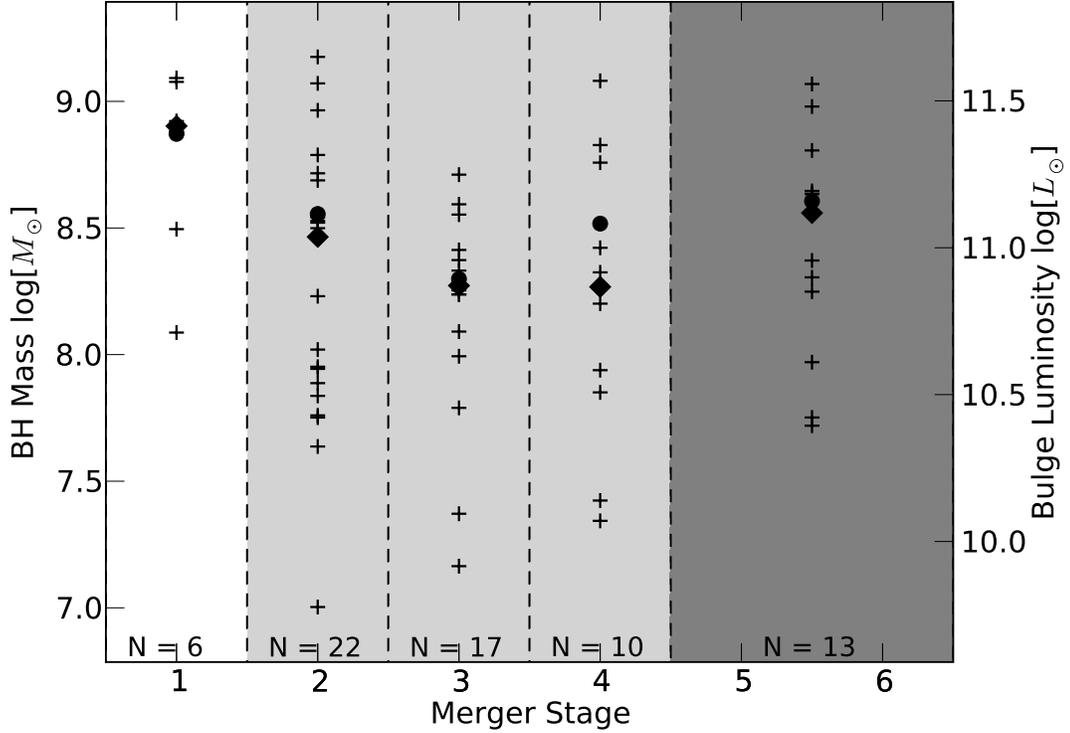}
\end{center}
\caption{The distribution (cross marker) of the central black hole masses (left label) along the merger stage sequence. The black hole masses are derived from the H-band bulge luminosities (right label) using the relation of \cite{Mar03}. The median and mean values of the central black hole mass for each bin are marked with a diamond and circle, respectively.  The stages (see text for details of the merger classification scheme) can be broadly characterized into pre-merging (1 in white), merging (2,3, and 4 in light grey), and post-merging galaxies (5 and 6 in dark grey; the nuclei have merged but the structure in the disk and tidal tails indicate the source has gone through a merger). Note that we have combined merger class  5 and 6 as post-merger stage LIRGs.}
\label{bhmass_stage}
\end{figure}

\begin{figure}[ht]
\begin{center}
\includegraphics[scale=0.7]{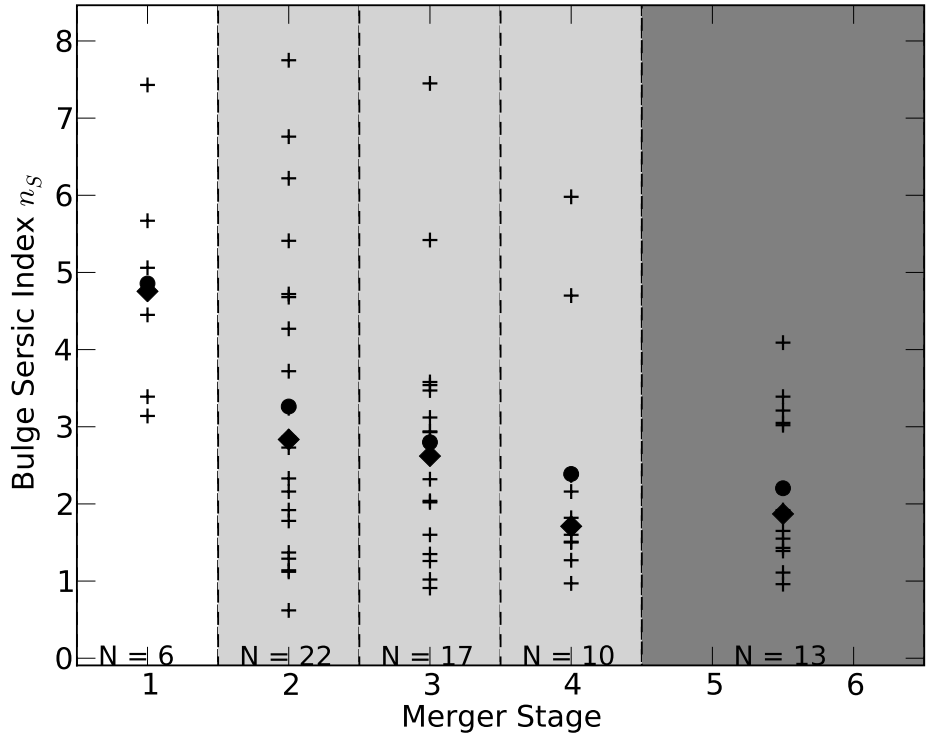}
\end{center}
\caption{The bulge S\'{e}rsic index $n_s$ along the merger stage sequence. The S\'{e}rsic index defines the degree of curvature of the radial brightness profile (see Equation~\ref{eq:sersic}). A smaller value of $n_s$ indicates a less centrally concentrated profile and a shallower (steeper) logarithmic slope at small (large) radii. Merger classification and markers are defined as in Fig.~\ref{bhmass_stage}. }
\label{sersic_stage}
\end{figure}

\begin{figure}[ht]
\begin{center}
\includegraphics[scale=0.7]{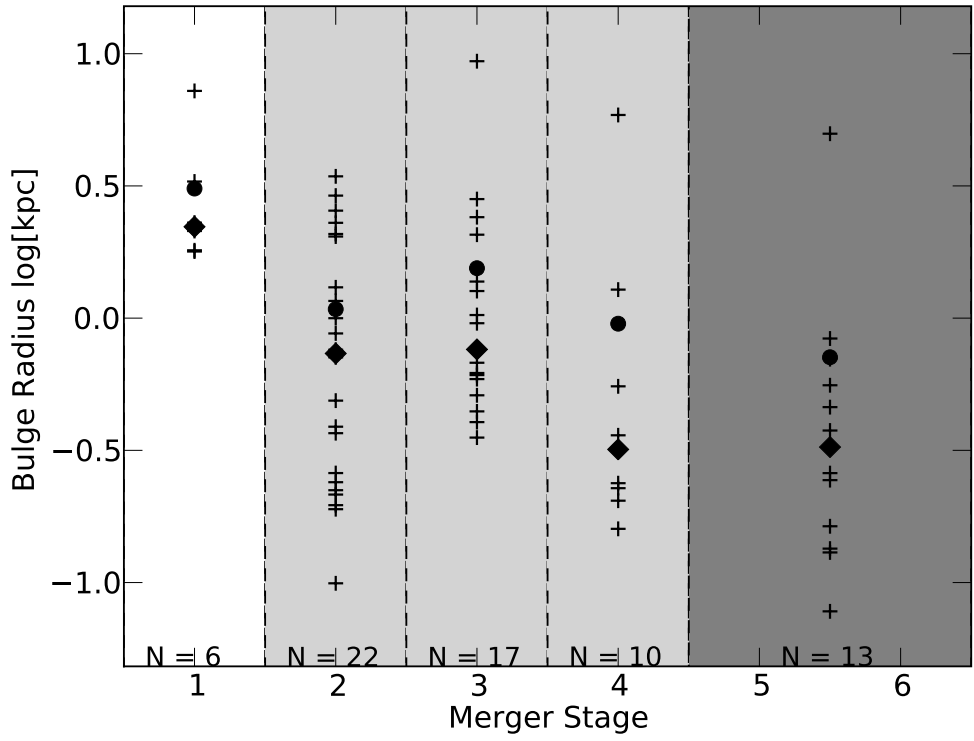}
\end{center}
\caption{The bulge radius $R_{Bulge}$ along the merger stage sequence. Merger classification and markers are defined as in Fig.~\ref{bhmass_stage}. }
\label{rad_stage}
\end{figure}

\begin{figure}[ht]
\begin{center}
\includegraphics[scale=0.7]{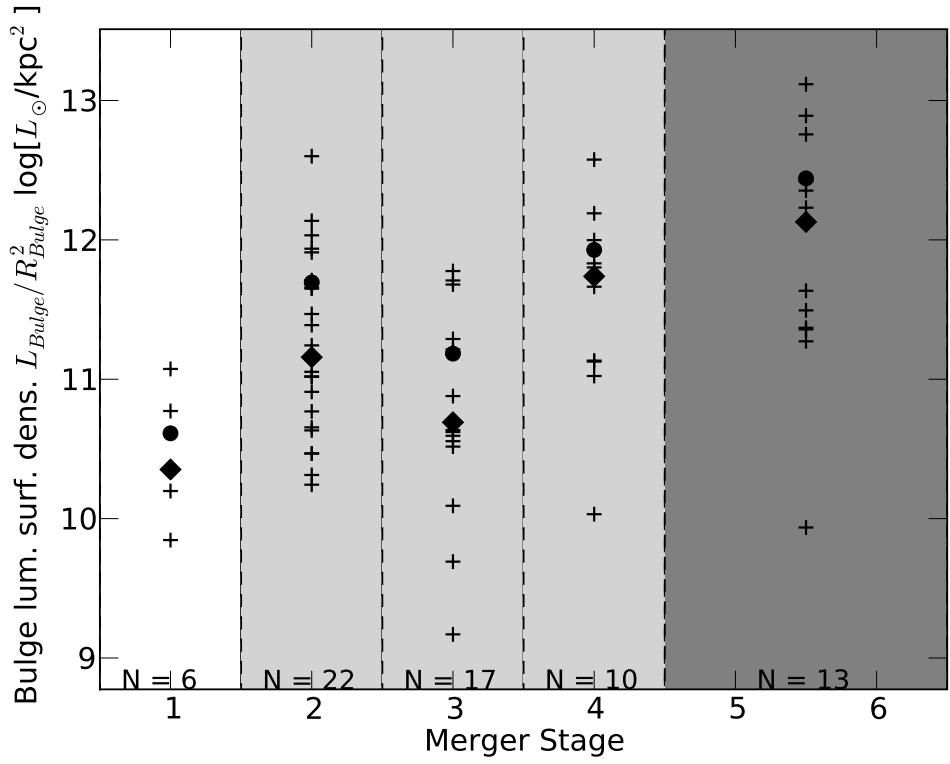}
\end{center}
\caption{The bulge luminosity surface density as defined as $L_{Bulge}/R_{Bulge}^2$ along the merger stage sequence. Merger classification and markers are defined as in Fig.~\ref{bhmass_stage}. }
\label{dens_stage}
\end{figure}

\begin{figure}[ht]
\begin{center}
\includegraphics[scale=0.38]{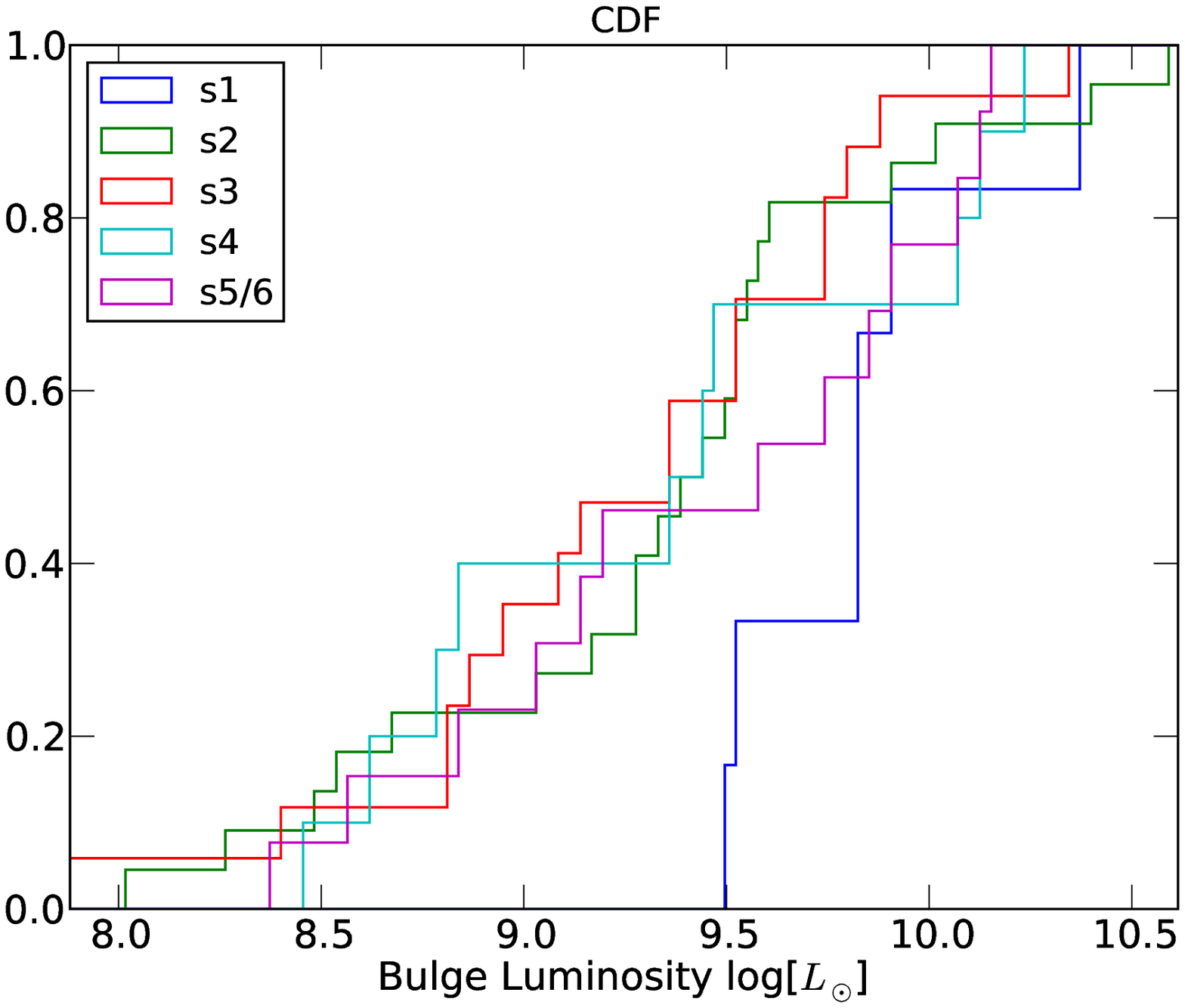}
\includegraphics[scale=0.38]{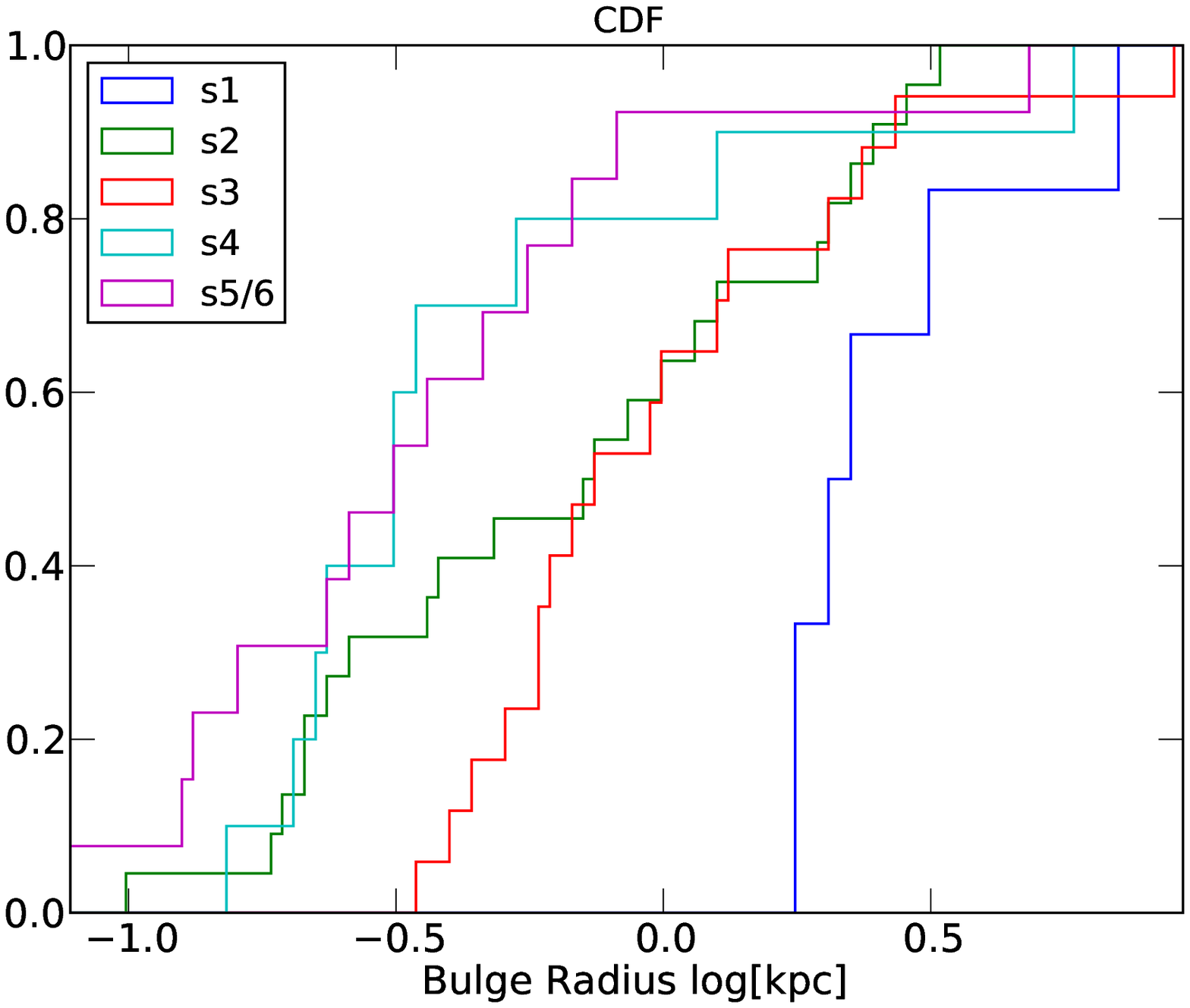}
\includegraphics[scale=0.38]{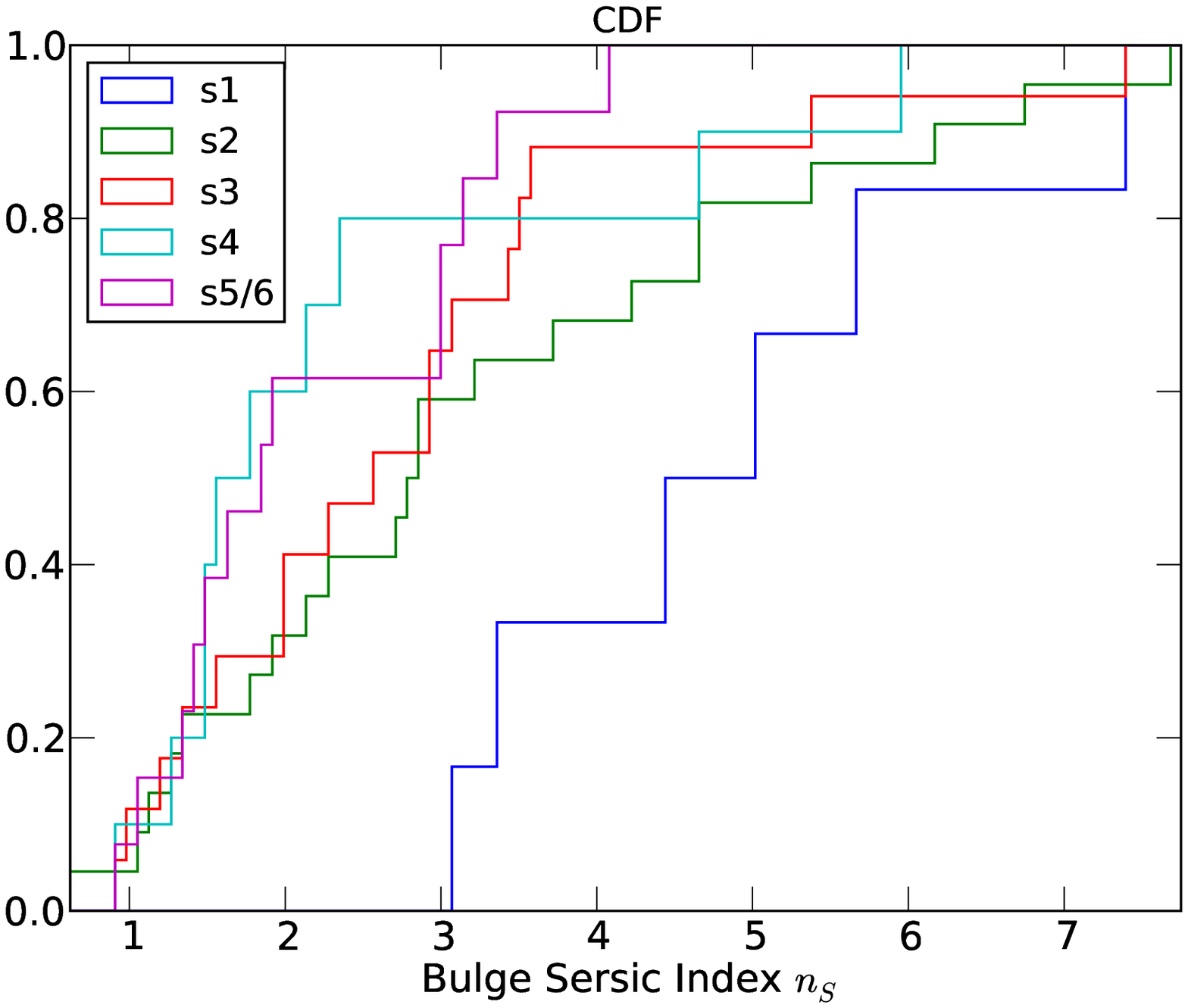}
\includegraphics[scale=0.38]{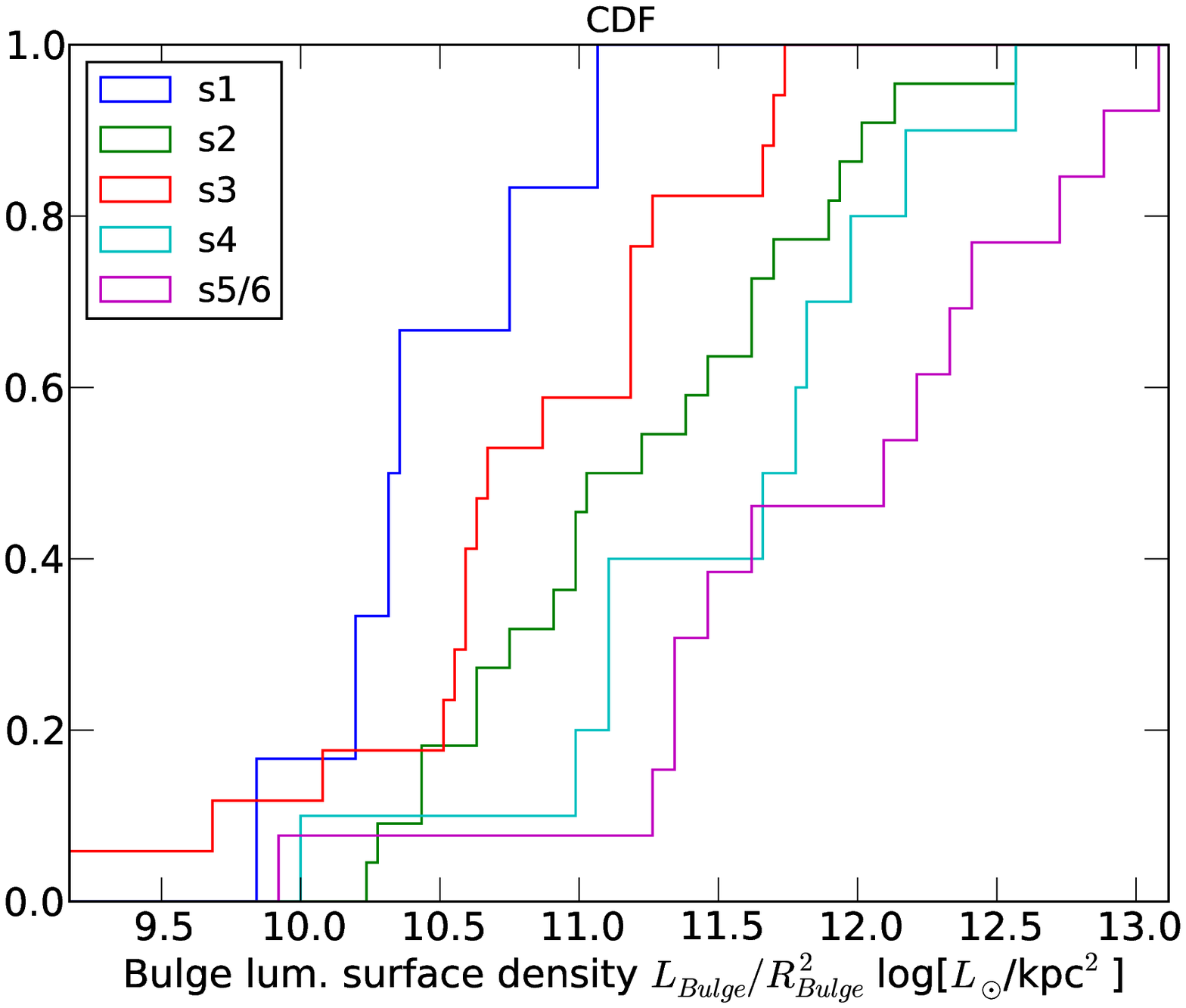}
\end{center}
\caption{The Cumulative Distribution Functions (CDF) of the bulge luminosity (top left), effective bulge radius (top right), bulge S\'{e}rsic index (bottom left), and bulge luminosity surface density (bottom right) for each merger stage (legend; see text for detailed description of merger stage classification). Note that we have combined merger class  5 and 6 as post-merger stage LIRGs.}
\label{cdf}
\end{figure}

To investigate the effect of the merger process on the central evolution of galaxies, such as the growth of the central black hole (BH) or stellar concentration, we study several morphological parameters as a function of merger stage: BH masses and bulge luminosities (Fig.~\ref{bhmass_stage}), bulge S\'{e}rsic index (Fig.~\ref{sersic_stage}), bulge radius (Fig.~\ref{rad_stage}), and the bulge luminosity surface density (Fig.~\ref{dens_stage}). An overview of the  mean, median, mean systematical error (from GALFIT), and the standard deviation of the bulge properties for all merger stages is presented in Table~\ref{tab:stages}.  To compare the bulge properties between different merger stage populations we performed Kolmogorov-Smirnov (KS) and Mann-Whitney U (MWU) tests (presented in Table~\ref{tab:ks_test}) which are based on the Cumulative Distribution Functions (CDF) of the bulge properties for each merger stage (see Fig.~\ref{cdf}). \par  

As shown in Fig.~\ref{bhmass_stage}, the BH masses (bulge luminosities) show an increase from merger stage 3 to merger stage 4--6 by a factor of 1.5--1.8 in their mean and median values. The KS and MWU test results confirm that the distributions of the values in each of these bins differ from each other (at a probability level of $\sim$90\%, see results in Table~\ref{tab:ks_test}). The bulge S\'{e}rsic index (see  Fig.~\ref{sersic_stage}) tends to decrease from merger stage 1 (mean $n_s=4.8$) to 5 (mean $n_s=2.2$). However, a significant change of the distribution of the bulge S\'{e}rsic index along the merger stage populations 2--6 is not supported by the KS and MWU tests. We find that the bulge radius $R_{Bulge}$ decreases significantly along all merger  stages (shown in Fig.~\ref{rad_stage} and Table~\ref{tab:ks_test}) from a median radius of 2.2 (merger stage 1) over 0.8~kpc (merger stage 2--3) to 0.3~kpc (merger stage 4--6). With decreasing bulge radius, the bulge luminosity surface density $L_{Bulge}/R_{Bulge}^2$ increases about a factor of 5--60 (median values) along the merger sequence (1--6, except 3) as shown in Fig~\ref{dens_stage} and statistically verified in Table~\ref{tab:ks_test}.\par

Interestingly, the LIRGs in our sample that show no interaction features (single isolated galaxies and merger stage 1 galaxies), have on average a significantly larger bulge luminosity (by a factor of $\sim$2), bulge radius (a factor of $\sim$1.5), bulge S\'{e}rsic index (a factor of $\sim$1.5), and a smaller luminosity surface density (a factor of $\sim$10) than LIRGs in merger stage 2--6. This is also confirmed by the KS and MWU test results and we will discuss possible reasons in \S~\ref{subsec:evo}.\par

\begin{figure}[ht]
\begin{center}
\includegraphics[scale=0.7]{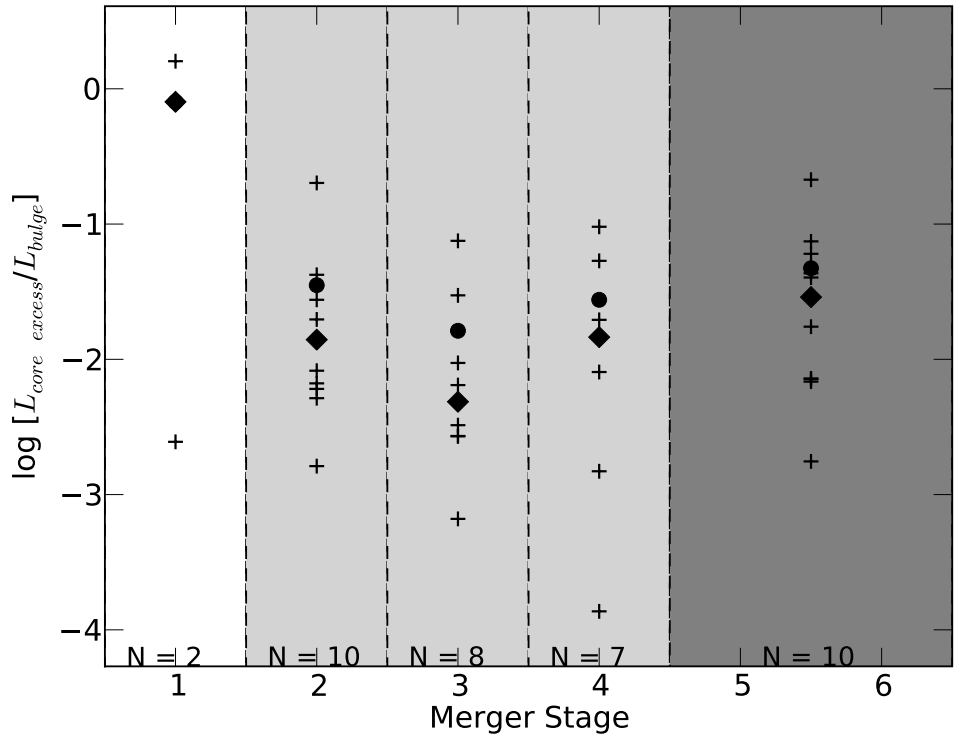}
\end{center}
\caption{The core excess luminosity fraction defined as the ratio of the core excess luminosity to bulge luminosity $L_{excess}/L_{Bulge}$ along the merger stage sequence. Merger classification and markers are defined as in Fig.~\ref{bhmass_stage}. }
\label{res_stage}
\end{figure}

We find that $\sim$20\% of our sample exhibit a core excess to total bulge luminosity $L_\mathrm{excess}/L_\mathrm{Bulge}>0.01$, but there is no clear correlation between $L_\mathrm{excess}/L_\mathrm{Bulge}$ and merger stage  (see Fig.~\ref{res_stage}). We find that two non-interacting LIRGs (AM~0702-601 North and IRAS~19542+1110) have a very large core excess to total bulge luminosity $L_\mathrm{excess}/L_\mathrm{Bulge}$  which is seen as a very bright central cusp in the center of these galaxies. A possible contribution of AGN activity to the core excess light is discussed in \S~\ref{subsec:dis-AGN}.\par

\clearpage

\subsection{Non-Nuclear Mid-IR emission}
\label{subsec:SF_outside}

\begin{figure}[ht]
\begin{center}
\includegraphics[scale=1.0]{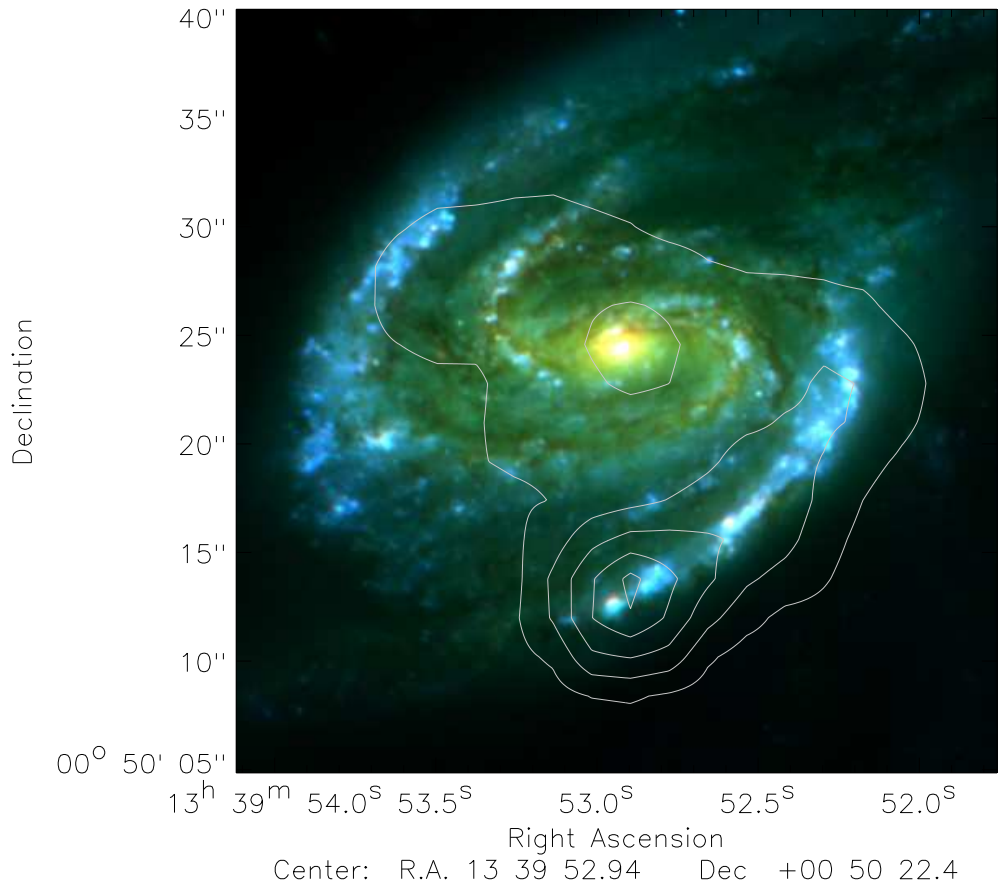}
\end{center}
\caption{NGC~5257; three-color composite image (red: H-band, green: I-band, blue: B-band) with MIPS 24~$\mu$m contours.}
\label{NGC5257}
\end{figure}

Although most of the LIRGs have peaks in their 24$\mu$m MIPS emission that correspond to peaks in their H-band NICMOS emission (i.e. the stellar nuclei), there are a few LIRGs with warm dust emission that is clearly offset from the stellar nucleus. A typical example for such a galaxy is II-Zw-096 which clearly shows that 80\% of its total infrared luminosity comes from an extremely compact red source not obviously associated with the nuclei of the merging galaxies \citep[see][]{Ina10}. Other similar galaxies in our sample are IC~1623 ($>$80\% of MIPS-24$\mu$m from a secondary nucleus), ESO~593-IG008 ($>$80\% of 24$\mu$m originates from possible secondary nucleus), and IRAS~F14348-1447 (24$\mu$m emission peak between double nuclei). One possible explanation is that these galaxies are presumably not yet relaxed with off-nuclear starbursts and/or strong shocks. 
Moreover, three LIRGs show also dominant off-nuclear 24$\mu$m emission that is very likely associated with extended star formation in spirals arms: Arp~256 North ($>$70\% of 24$\mu$m in spiral arm), NGC~5257 ($>$80\% from outer spiral arm), and NGC~6670 West ($\sim$50\% from spiral arm). In particular one LIRG double system, namely NGC~5257, stands out from the rest of our sample, as it shows that almost all of its mid-IR emission is generated at the end of one of the spiral arms (in both galaxies), $\sim$7~kpc away from the center (Fig.~\ref{NGC5257}, note that only one galaxy is shown due to the limited FOV of NICMOS). This seems to be very surprising given their very regular spiral disks and that no visible counterpart is seen at the center of the mid-IR emission, neither in the H- nor in the B-band image. This phenomenon might be also associated with Tidal Dwarf Galaxy (TDG) formation, which is also suggested for other interacting galaxies \citep{Duc98, Duc00}. Since NGC~5257 consists of two galaxies with a projected separation distance of $\sim$50 kpc, the exact cause of the enhanced disk emission is not clear. Similar cases are well known for, e.g., the Antennae galaxies \citep{Mir98} and the LIRG Arp299 \citep{Cha02, Char04}.

\clearpage

\section{Discussion}
\label{sec:dis}

\subsection{Obscured Nuclei}
Understanding the exact number of merger progenitors for LIRGs is important as these numbers provide information about merger timescales and subsequently LIRG activity timescales \citep[see][]{Mur96}. 
The study of the small and large-scale morphologies of LIRGs in the nearby universe allows us to understand the physical processes that drive galaxy evolution. While ULIRGs are predominantly merger systems, the fraction of merger systems and the merger timescales for LIRGs are more controversial: The imaging analysis of 30 local LIRGs \cite{Alo06} indicates that most of these LIRGs (log[L$_{IR}$/L$_\odot$]$= 11.0-11.9$) have prominent spiral patterns, and of those, a non negligible number are weakly interacting or even isolated systems. \cite{Bri07} find in an analysis of MIPS 24~$\mu$m detected and undetected mergers ($0.2 \leq z \leq 1.3$), that a larger fraction of LIRGs appear later, in the merger phase, than in the pre-merger, close pair phase. 
However, it is expected that the fraction of major mergers increases with IR luminosity. Thus far, the trigger mechanisms and end products of LIRGs have not been constrained as well as those of local ULIRGs.\par

We find for our sample that the fraction of LIRG systems (log[L$_{IR}$/L$_\odot$]$= 11.4-12.5$) with at least two interacting nuclei is 63\% (see Fig.~\ref{nuclei_H-B}). Note that the fraction of major merger for the entire GOALS sample (log[L$_{IR}$/L$_\odot$]$>11$) is $\sim$50\% (based on Spitzer IRAC and HST data). Interestingly, the comparison with our HST ACS B-band images revealed that roughly half of the double nuclei seen in the HST NICMOS H-band are obscured by dust (not visible in the B-band). Therefore, NIR observations of LIRGs and ULIRGs at high-redshift (z$>$2), which correspond to the B-band rest-frame range, might require significant correction factors of $\sim$2 or higher to accurately estimate the true number of multiple nuclei systems.\par 

In particular, \citet[using HST NICMOS F160W]{Das08} found that the fraction of binary systems in ULIRGs at redshift 2 is roughly 50\%, which is significantly smaller than that of the local ULIRG population \citep[nearly 100\% of objects in the IRAS 1 Jy sample show signs of major merger interaction][]{Kim02, Vei02}. Also in studies of \citet[MIPS 24~$\mu m$ and HST ACS F850LP filter]{Bel05} and \citet[MIPS 24~$\mu m$ and HST ACS Viz band]{Mel05} major merger appear to make up a smaller fraction of z$\simeq0.7$ LIRGs compared to local systems. The strong wavelength dependence on the number of visible nuclei may explain some of the apparent discrepancy between the (U)LIRG population at local and high-redshift, particularly in identifying late stage mergers when tidal tails have faded and double nuclei are surrounded by dust. 

\subsection{Merger Time Scales}
The space density of LIRGs combined with an estimate of the time-scales of their power sources (star formation and/or AGNs) provides important constraints for galaxy evolution scenarios that link the IR-luminous phase with the luminous, unobscured AGNs observed in Quasar Hosts \citep{San88b}. 
The merger time scale depends on the separation between the multiple nuclei observed in our NIR images (Fig.~\ref{hist_dist}) and we follow the approach by \cite{Mur96}, namely: In a first
stage we adopt a constant velocity $v_r$  until the nuclei are 10~kpc apart. Because of the small range in nuclear separations, and the large uncertainties involved, we treat the radial velocity as constant. In this case, the time from observation until the nuclei are within 10~kpc is ($r-10$~kpc)/$v_r$ with $r$ as the projected nuclear separation. The time spent in the second stage is estimated by calculating the dynamical time scale times the mass ratio of the two nuclei, given as $(M_1/M_2)\cdot 2\pi r/v_c$ with the mass ratio $M_1/M_2$, the projected nuclear separation $r$ and the orbital velocity $v_c$ \citep[see][]{Mur96, Bin08}.  As the nuclei spiral together, the relative velocity is similar to the circular velocity at that radius \citep{Bin08}. In the calculations that follow we adopt an average radial velocity $v_r$ and an average circular velocity $v_c$, of  200~km~s$^{-1}$ for all the double nucleus galaxies in our sample \citep{Mur96}.
These average velocities are supported by simulations \citep[e.g. see Fig.~15 in][]{Xue08} and the distribution of  the line-of-sight velocities differences between the galaxy pairs in our sample (typically between 60--180~km~s$^{-1}$; note that the radial velocity is on average a factor of 1.5 larger than the line-of-sight velocity component). 
It is important to note that for an individual galaxy all we can estimate from the images and these simple dynamical arguments is the time from the present until the time of final merger, when the nuclei of interacting galaxies coalesce. We must rely on the distribution of these time scales within the sample, along with the morphologies, to allow a calculation of the average remaining merger time scales of our sample.  
For our LIRG sample (log[L$_{IR}$/L$_\odot$]$>11.4$) we find a median remaining time until the nuclei merge ($t-t_m$) of $4.3 \cdot 10^8$~yrs. 

\begin{figure}[ht]
\begin{center}
\includegraphics[scale=0.6]{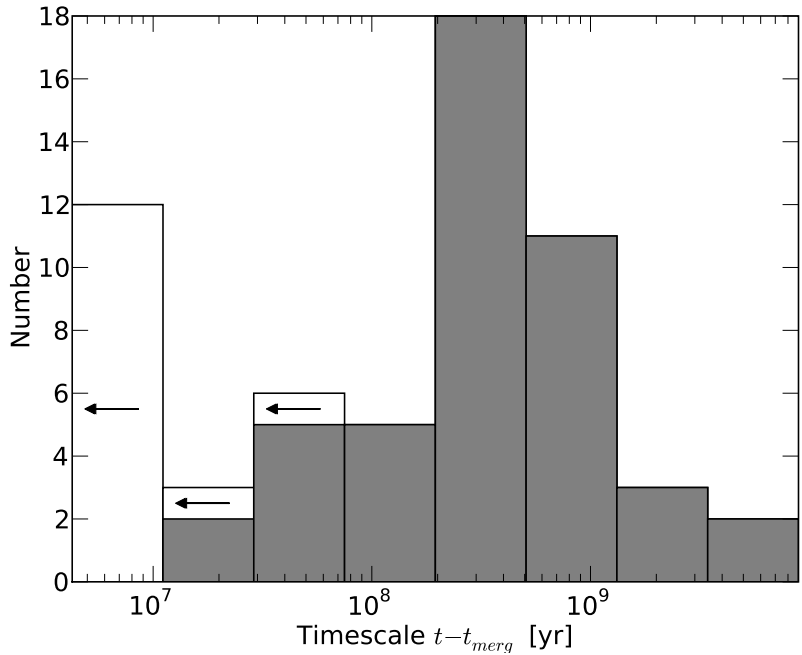}
\end{center}
\caption{Histogram of the remaining merger time scale, $t - t_{merg}$ (see text for details) with a median time scale of $4.2 \cdot 10^8$~yrs. Fourteen apparent single nuclei in galaxies that exhibit prominent interaction features, such as tidal tails, are also included in white, converting our angular resolution limit of  $0.15\arcsec$ to the linear size at their corresponding distance and assuming a nuclei mass ratio of 1:1.}
\label{hist_time}
\end{figure}

\begin{figure}[h]
\begin{center}
\includegraphics[scale=0.7]{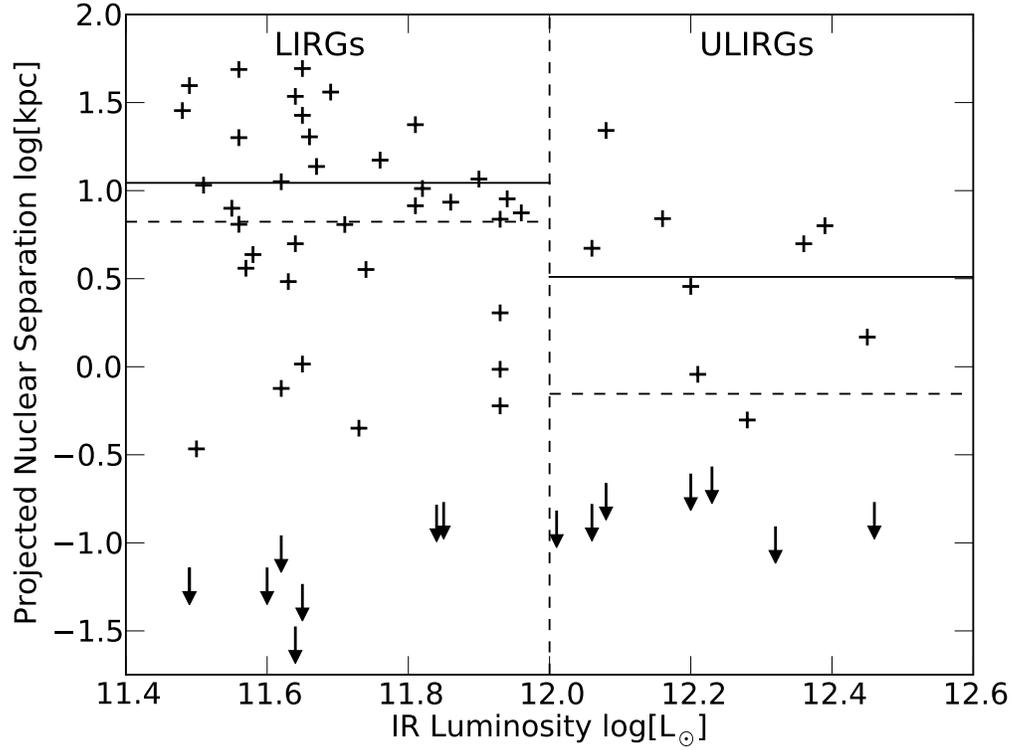}
\end{center}
\caption{The projected nuclear separation as a function of IR luminosity for the observed double nuclei (plus marker), as well as apparent single nuclei (arrows down, converting our angular resolution limit of  $0.15\arcsec$ to the linear size at their corresponding distance) whose host galaxies exhibit interaction features. The mean and median values of the projected nuclear separation are shown as solid and dashed horizontal lines, respectively.}
\label{IR_dist}
\end{figure}

As shown in Fig.~\ref{hist_time}, we find two peaks in the numbers of LIRGs as function of the remaining merger time scale: The largest occurs at 0.3~Gyr$< [t - t_{merg}] < $1.3~Gyr, roughly representing the first passage of interacting galaxies, while the second peak occurs at the final coalescence of the nuclei (most at $[t - t_{merg}] < 10^7$~yr), including all galaxies with apparent single nuclei that show interaction features.  Note that the second peak represents the same number of LIRGs and ULIRGs (each 6). The fact that we see two peaks is consistent with merger simulations \citep{Cox08, Hop08b}, which show two star formation rate (SFR) peaks as function of merger time.  Furthermore, the time-scales of these two peaks are roughly consistent with the two SFR peaks as seen the in merger simulations, but we see slightly more LIRGs at  $[t - t_{merg}] < 1$~Gyr. In detail, the peak at $[t - t_{merg}] \sim 1$~Gyr  is a bit broader (0.3~Gyr$< [t - t_{merg}] < $1.3~Gyr) than we would expect from the merger models  (0.8~Gyr$< [t - t_{merg}] < $1.3~Gyr), indicating that we observe more LIRGs at shorter timescales. The fraction of interacting LIRGs is larger for the peak at $\sim$1~Gyr (53\% of interacting LIRGs) than for the peak at the nuclear coalescence (26\%), which likely corresponds to the broader width  ($\sim$0.7 Gyr) of the SFR peak at $\sim$1~Gyr than the narrow peak ($\sim$0.3 Gyr) at the nuclear coalescence \citep[see][]{Cox08, Hop08b}. As our GOALS sample is a flux-limited snapshot of galaxies passing through the LIRG and ULIRG phase, the data suggest that galaxies spend a longer time in the LIRG phase, which is more easily associated with the first peak in the model SFR at $[t - t_{merg}] < 1$~Gyr. Furthermore, the IR luminosity is slightly larger for LIRGs at  $[t - t_{merg}] \simeq 0$ (median log[$L_{IR}/L_{\odot}] = 11.7$) than for LIRGs at $[t - t_{merg}] \simeq 1$ (median log[$L_{IR}/L_{\odot}] = 11.9$). This is consistent with our finding that LIRGs in late stages of merging have significantly larger total IR luminosities (roughly a factor of two) than pre- or non-merging LIRGs. We also discuss this point below (see Fig.~\ref{IR_dist} as well). \par

To estimate the timescale that LIRGs spend in the final nuclear coalescence (post-merger stage), we have applied the following approach: In principle, the post-merger time scale can be estimated by multiplying the merger time-scale with the ratio of galaxies with single nuclei to double nuclei. However, this approach might be oversimplified since not all LIRGs might have undergone a merging process, and instead other mechanisms may be responsible for a significant fraction of LIRGs. In fact, 23\% of LIRGs in our HST NICMOS sample show no major interaction features in the H-, I-, or B-band (merger stage 0 and 1, see \S~\ref{subsec:res-GALFIT}) and hence it is unlikely that they have undergone a previous major merger. Therefore, we take into account only galaxies that are classified as mergers (merger stage 2---6, see \S~\ref{subsec:res-GALFIT}). The time spent in the pre-merger stage for a random galaxy is statistically roughly twice the average observed time required to complete the merging $t-t_m$ \citep[see][]{Mur96}, or about 10$^9$yrs for our sample. Thus, the time spent in the post-merger stage is the median merger time of LIRGs with double nuclei ($2 \times 0.43$ Gyr) times the ratio of the number of single nuclei with interaction features (15 LIRGs) to double nuclei (45 LIRGs). This results in an average post-merger time of $\sim 2.9\cdot 10^8$~yrs, which is very similar to the typical timescale of the SF burst of 0.3 Gyr as seen in major merger simulations \citep{Cox08, Hop08b}.\par

Since we expect most ULIRGs to pass through a LIRG phase, we might expect to see a correlation of merger stage, nuclear separation and IR luminosity \citep{Mur01, Vei06}.  A comparison of LIRGs and ULIRGs in our HST NICMOS sample, presented in Fig.~\ref{IR_dist}, reveals that the projected nuclear separation is significantly different between both populations (probability level of 98\% in KS test) with a mean (median) projected separation of 3.4~kpc (1.2~kpc) and 11.1~kpc (6.7~kpc) for ULIRGs and LIRGs, respectively. \par 

The results for our subsample of ULIRGs (log[$L_{IR}/L_{\odot}]>12.0$) are also in good agreement with a study for a larger sample of ULIRGs \citep{Mur96}, which is based on ground-based observations (seeing of $\Delta R \simeq 0.8\arcsec$ for K band images), and hence the lower spatial resolution limits the ability to resolve double nuclei. Including galaxies with apparent single nuclei (upper limits on the separation of double nuclei), most of the ULIRGs in the sample of \cite{Mur96} (more than 60\%) have nuclear separation between 1-4~kpc (median of $\sim$1.8~kpc) which is very similar to our results for the ULIRG population (median of $\sim$1.2~kpc). Note that we have an overlap of ten ULIRGs with the sample of \cite{Mur96} and we found for one ULIRG, namely IRAS~F19297-0406, two nuclei with a projected nuclear separation of $0.77\arcsec$ that was previously characterized as an apparent single nucleus by \cite{Mur96} due to limited spatial resolution of the R-band image (1.9\arcsec). However, only our entire sample covering a IR luminosity range of log[$L_{IR}/L_{\odot}]>11.4$, allows us to see a strong decrease of the projected nuclear separation as a function of IR luminosity, as shown in Fig.~\ref{IR_dist}. 

\clearpage

\subsection{The Evolution of the Central Stellar Structure along the Merger Stage}
\label{subsec:evo}
 
Evidence suggests that ULIRGs may evolve into elliptical galaxies once the starburst subsides and the gas is either used up or expelled in a wind \citep{San88b, Hec90, Gen01, Vei06}. Therefore, the stellar surface brightness profiles of LIRGs should provide a glimpse of the process of bulge and black hole building. Unlike previous studies, which focused primarily on ULIRGs (log[$L_{IR}/L_{\odot}]>12.0$), the GOALS NICMOS sample targets the nuclear regions of all merger systems in the IRAS Bright Galaxy Sample \citep[RGBS;][]{San03} with log[$L_{IR}/L_{\odot}]>11.4$. Because of the low red-shift range of our sample ($0.01 < z < 0.05$), the galaxies are bright and well-resolved due to their large angular size.\par

Although we find a significant increase in the bulge luminosity surface density (a factor of 5 to 60) along the merger sequence, the growth of the BH mass (bulge luminosity) toward later merger stages is only marginal: A factor of $\sim1.8$ in bulge luminosity from merger stage 3 to 5/6 with a probability level of $\sim$90\% that both merger stage populations are not drawn from the same parent population (KS and MWU test, see Table~\ref{tab:ks_test}).
Two possibilities may explain why we do not see a a more significant increase of the bulge luminosity: First, as the merging proceeds, the bulge may get partially disrupted or stripped and may assemble again after the LIRG phase has passed. Thus, our measured bulge luminosities may miss a significant bulge fraction at the late merger stage, resulting in smaller estimated black hole masses. A simpler explanation could be that the flux limited nature of our HST LIRG sample (log[$L_{IR}/L_{\odot}]>11.4$) excludes galaxies with smaller bulge masses in single isolated galaxies and galaxies in the pre-merging stage. Indeed, we find evidence for this, as the LIRGs in our sample that show no evidence for interaction features (single isolated galaxies and merger class 1) have on average a significantly larger BH mass (a factor of two). Since LIRGs are drawn from the top end of the IR luminosity function and mergers boost IR luminosity, we might expect the non-interacting LIRGs to be more luminous and hence more massive than the progenitors of interacting LIRGs. The interacting LIRGs would tend to include more systems with smaller bulges and then grow their bulges during the merging process. If this is the case, the non-interacting LIRGs would tend to have more massive bulges - exactly as we observed in our sample. A detailed comparison of the bulge masses of non-interacting spiral galaxies with log[L$_{IR}$/L$_\odot$]$ < 11.4$ would be needed to test this hypothesis.

\subsection{Merger-induced Building of Central Starbursts: Comparison to Models}
\label{subsec:model}
Dissipation in mergers can generate central starbursts, imprinting a central ``extra light'' component into the surface brightness profiles of merger remnants \citep{Her93}.
Such an excess of the central light or ``cuspiness'' in the surface brightness profiles has been found for some elliptical galaxies \citep{Fer06, Cot07, Kor09, Hop09a}. This can be explained by the assumption that the envelopes of cusp ellipticals are established by violent relaxation in mergers acting on stars present in gas-rich progenitor disks, while their centers are structured by the relics of dissipational, compact starbursts \citep{Hop09b}. Given the evidence that ULIRGs likely evolve into massive elliptical and S0 galaxies \citep{Gen01, Tac02} and the fact that LIRGs and ULIRGs host powerful starbursts, the study of our LIRG sample as a function of merger stage allows us in principle to test the build-up process of nuclear starbursts. \par

To test such a scenario we studied the following parameters as a function of merger stage (1---6, see \S.~\ref{subsec:res-GALFIT}): (1) the effective bulge radius $R_{Bulge}$, (2) the bulge luminosity $L_{Bulge}$, (3) the bulge S\'{e}rsic index $n_s$ which defines the steepness of the inner bulge profile (see Fig.~\ref{sersic_stage}), (4) the bulge surface brightness defined as $L_{Bulge}/R^2_{Bulge}$, and (5) the central light concentration given by the ratio of core excess luminosity to total bulge luminosity (see Fig.~\ref{res_stage}). We have compared our results with recent models of \cite{Hop09a}, which are based on N-body simulations and smoothed particle hydrodynamics, account for radiative cooling and for heating by a UV background, and incorporate a subresolution model of a multiphase interstellar medium (ISM) to describe star formation and supernova feedback. Since the model of \cite{Hop09a} focuses on local gas-rich mergers (unlike high-redshift simulations which have a factor of 2--5 larger gas content), it is ideally suited for a comparison to our data-set of local major mergers.

\subsubsection{Bulge Radius and Surface Brightness}
The merger model of \cite{Hop09a} predicts a decrease of the effective bulge radius along the merger process due to gas inflow: Tidal torques excited during major merger provide the fuel to power intense starbursts boosting the concentration and central phase-space density. As shown in \cite{Hop09a} a gas inflow of e.g $\sim$10\% shrinks the effective bulge radius to about half its previous size and subsequently enhances the bulge surface brightness. Our findings show the same trend as the model predictions (see Fig.~\ref{rad_stage} and Fig.~\ref{dens_stage}), given a decrease (increase) of the effective bulge radius (surface brightness) by a factor of $\sim 2-3$ ($5-20$), and hence provide evidence for the idea that gas inflow can dramatically reduce the apparent bulge radius during a merger.
  
\subsubsection{Bulge S\'{e}rsic Index}
Due to the violent relaxation of stars during a merger, the model of \cite{Hop08a} predicts an increase of the bulge S\'{e}rsic index $n_s$ towards late merger-stage (from $t - t_{merg}=0.5$~Gyr to $t - t_{merg}=0$~Gyr). We do not find a significant increase of $n_s$ (see Fig.~\ref{sersic_stage} and Table~\ref{tab:ks_test}). This discrepancy can have several reasons, such as that young stellar populations and age (metallicity) gradients can shift $n_s$ significantly and/or tidal interactions are changing $n_s$ during the more active/merging phases. 

\subsection{The Role of AGN Activity and Young Stellar Populations on Nuclear Properties}
\label{subsec:dis-AGN}

Throughout most of this paper, we have assumed that the near-infrared H-band data is an accurate tracer of the stellar mass, as it is less affected by dust extinction (than the UV or optical), and has less of a contribution from young, blue stars.  However, LIRGs and ULIRGs are known to be powered by starbursts and AGN, which can, in some cases, significantly contribute to the near-infrared light. In the following sub-sections we discuss the possible contributions of central starbursts and AGN to the near-infrared light.

\subsubsection{Young Stellar Populations} 
Evidence suggests that asymptotic giant branch (AGB) and red supergiant (RSG) stars can dominate the NIR in the center of some starburst galaxies \citep{Arm95, Rot10, Mel10}. Since high-spatial resolution spectroscopy is required to identify and resolve the young stellar populations, their spatial distribution is unknown for our sample. However, two scenarios can constrain the possible effects of young stellar populations on our measured nuclear properties: (1) If young stars form in the core of the galaxies at scales $\lesssim$ 300~pc, these populations would be taken into account by our measured PSF component and hence not affect our measured bulge properties. (2) In case young stars are more widely distributed on scales typical for stellar bulges (median effective bulge radius for our sample is 0.7~kpc), they could affect our measured properties. If young stars form in a disk, they would manifest an exponential radial profile, and a central GALFIT component with a S\'{e}rsic index $n_s\simeq1$ (exponential disk) would be expected, instead of $n_s>1$ (more typical for old stellar bulges), which we find for most of our galaxies (see Fig.~\ref{rad_stage}). If this is not the case,  namely that the majority of young stars form at bulge scales ($>300$~pc, not in in the core), are distributed in a spheroid rather than in a disk, and dominate the NIR, then the H-band bulge luminosity may not be a reliable estimate of the central BH mass. However, this does not affect our comparison of the central stellar structure along the merger stage to merger models (\S.~\ref{subsec:model}), since young stellar contributions are already taken into account in the merger simulations  \citep{Hop09a, Hop09b}.

\subsubsection{AGN Activity}
As described in \S~\ref{subsec:res-GALFIT} we find for some galaxies a large ratio of core excess to total bulge luminosity $L_\mathrm{excess}/L_\mathrm{Bulge}$. One possible explanation of such a bright central light emission could be that the LIRG activity is caused by AGN heating. In general, lower Equivalent Widths (EQWs) of the 6.2~$\mu$m PAH feature are associated with AGN activity \citep[see e.g.][]{Gen98, Stu00, Arm07, Des07} because the hot dust continuum increases and the hard AGN photons may destroy/ionize the PAH molecules. \cite{Pet10} estimated that AGN are responsible for $\sim$12\% of the total bolometric luminosity of local LIRGs based on several mid-IR line diagnostics measured with the Infrared Spectrograph on Spitzer. \par 

\begin{figure}[ht]
\begin{center}
\includegraphics[scale=0.7]{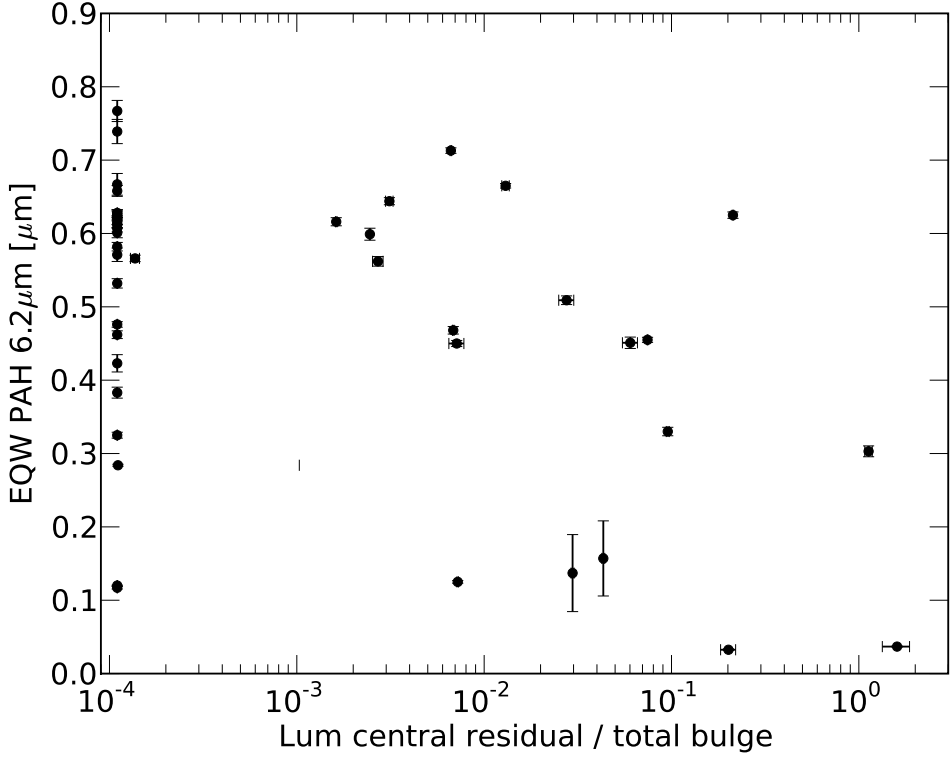}
\end{center}
\caption{Equivalent Width (EQW) of the 6.2~$\mu$m PAH feature as function of the core excess luminosity to total bulge luminosity. 
A smaller EQW indicates a larger contribution of AGN generated emission to the mid-IR emission. No significant trend is visible between 6.2~$\mu$m PAH EQW and core excess luminosity (see text).}
\label{EQW_res}
\end{figure}

We find no significant trend between the ratio of core excess to bulge luminosity and the PAH EQW (see Fig.~\ref{EQW_res}) of the 6.2~$\mu$m PAH feature \citep{Pet10}. However, one example of a galaxy where a central AGN might dominate the NIR emission is AM~0702-601: A galaxy in our sample with a very small PAH EQW (0.037~$\mu$m) but also exhibits one of the most significant core excess luminosity fractions ($L_\mathrm{excess}/L_\mathrm{Bulge} = 1.5$). Counter examples, where the core excess light fraction in the H band and PAH EQW are not correlated: ESO~060-IG016 and NGC~3690 East, which have very small PAH EQWs ($<$ 0.27~$\mu$m) and hence likely large AGN contribution, but they exhibit no significant core excess light fraction ($L_\mathrm{excess}/L_\mathrm{Bulge} < 0.001$). One possible explanation might be that those AGN systems are so deeply embedded in dust that they don't show a central light excess in the H-band, in contrast to some ULIRGs which show a very large light excess in the center when an AGN is present \citep[see e.g.][]{Sur99}.  
We find also galaxies (ESO~239-IG002, WKK~2031, and NGC~1614) with a significant core light excess ($L_\mathrm{excess}/L_\mathrm{Bulge} > 0.05$) but relatively large PAH EQWs (0.4 - 0.7~$\mu$m), suggesting that those galaxies build up a concentrated stellar ``cusp'' in the center due to luminous nuclear starbursts. The sources with the largest core excess ($L_\mathrm{excess}/L_\mathrm{Bulge} > 0.05$) are not responsible for the decrease in the bulge radius along the merger sequence, as the central PSF (even if it is due to an AGN) is already taken into account by the GALFIT fitting process (see \S.~\ref{subsec:model}).  We obtain roughly the same results for the bulge properties along the merger sequence if we exclude the galaxies with small 6.2~$\mu$m PAH EQW (EQW $<$ 0.27~$\mu$m) from our sample.\par

AGN activity might also effect the observed NIR luminosity and hence one would expect a possible contribution to the measured bulge luminosity. This could subsequently explain the increase of the BH mass (bulge luminosity) with larger IR luminosity (see Fig.~\ref{bulge_IRlum} and Eq.~\ref{eq:IR_bulge}). However, a possible contribution of AGN activity to the bulge luminosity is not very likely since most of our bulges are resolved and have typical ranges of 0.5-2~kpc while the main dust heating by AGN (500-2000K) is generated  on scales of (10-100)~pc \citep{Soi01}. Furthermore, the central light excess expected from an AGN contributes on average less than 5\% to the total bulge luminosity for our sample (see Fig.~\ref{EQW_res}).

\section{Summary}
We have studied the nuclear stellar properties of 73 LIRG systems (log[L$_{IR}$/L$_\odot$]$ = 11.4-12.5$) using high-resolution H-band images obtained with HST NICMOS revealing a large variety of double and triple nuclei merger systems. To investigate the effect of the merging process on the evolution of galaxies, such as the growth of the central BH or stellar concentration, we studied several morphological parameters as a function of merger stage using GALFIT. The main results are summarized as follows:
\begin{enumerate}
\item The fraction of LIRG systems with at least two interacting nuclei is 63\%. Triple nuclei systems are found for at least 3 (possibly 5) LIRG systems. The comparison of our NICMOS images with comparable resolution HST B-band images revealed that roughly 50\% of the double nuclei are obscured by dust, with 10 of the obscured nuclei responsible for the primary mid-IR emission (MIPS 24$\mu$m). This implies strong limitations on the ability to detect the true nuclear structures of luminous infrared galaxies at high-redshift and may explain some of the apparent discrepancy between the LIRG population at local and high-redshift \citep[e.g.][]{Kim02, Bel05, Mel05, Das08}.
\item ULIRGs (log[L$_{IR}$/L$_\odot$] $>$ 12.0) have significantly smaller nuclear separations than LIRGs (log[L$_{IR}$/L$_\odot$]$ =$ 11.4 --- 12.0) with a median value of 1.2~kpc and 6.7~kpc, respectively. In our sample, merger (regardless of whether LIRG or ULIRG) seem to be prevalent at two time scales (based on the projected nuclear separation and mass ratio of the nuclei): First, at a remaining merger time scale of  $0.3<[t-t_{merg}]<1.3$~Gyr (53\% of mergers in our sample), and second, at $[t-t_{merg}]\sim 0$ (26\%), likely representing the first passage of interacting galaxies and the final nuclear coalescence, respectively.  An average post-merger time (LIRG phase after the nuclei merged) of 0.3~Gyr is estimated. 

%The post-merger time that  the The number of mergers peaks at two The majority of mergers in our sample have remaining merger time scales (based on the projected nuclear separation and mass ratio of the nuclei) of $0.3<t-t_{merg}<1.3$~Gyr. 

%The median remaining merger time scale, $t-t_{merg}$, for mergers in our sample (based on the projected nuclear separation and mass ratio of the nuclei) is $4.3\times10^8$~yrs. The distribution of LIRGs along the remaining merger time shows two peaks, ione at $t-t_{merg}$
\item The bulge luminosity shows, on average, a small increase (about a factor of 1.8) towards late merger stages. Interestingly, the LIRGs in our sample that show no interaction features have on average a significantly larger bulge luminosity (a factor of two) than interacting LIRGs. This is likely to be a selection effect for more intrinsically luminous non-merging galaxies among the LIRG sample.
\item The bulge luminosity surface density $L_{Bulge}/R_{Bulge}^2$ increases significantly along the merger sequence which is primarily due to a decrease of the bulge radius towards late merger stages. These findings are in agreement with models that include gas inflow which can dramatically affect the measured bulge properties during the merging process. No significant dependence of the bulge S\'{e}rsic index is found as a function of merger stage.
\item Although no significant correlation between the 6.2$\mu$m PAH EQW and the core excess light in the H band is found, LIRGs with the largest core excess exhibit slightly smaller PAH EQWs, suggesting more hot dust emission and hence AGN activity. 
\item At least seven LIRGs in our sample exhibit warm dust emission ($>70$\% of its total infrared luminosity) that is clearly offset form the nucleus. Instead, their emission regions seem to be associated with spiral arms, a possible secondary nucleus, or the region between the merging nuclei.
\end{enumerate}

\par

We would like to thank Philip F. Hopkins for valuable discussions of our results in terms of model predictions and numerical simulations. This research has made use of the NASA/IPAC Extragalactic Database (NED) and Infrared Science Archive which is operated by the Jet Propulsion Laboratory, California Institute of Technology, under contract with the National Aeronautics and Space Administration. Support for this work was provided through grant HST~GO 11235.01-A by NASA from the Space Telescope Science Institute, which is operated by the Association of Universities for Research in Astronomy, Inc., under NASA contract NAS 5-26555.

\clearpage
\newpage

\begin{deluxetable}{lcclll}
\tabletypesize{\tiny}
\tablecaption{Sample Overview}
\tablehead{\colhead{Name}  & \colhead{RA} & \colhead{DEC} &  \colhead{log($L_{IR}/L_{\odot}$)}& \colhead{D} &  \colhead{HST ID} \\
       & (J2000) & (J2000) &  & [Mpc] & }
\startdata
NGC0034&00:11:06.61&-12:06:27.4&11.49&84.1&7268\\
ARP256N&00:18:50.12&-10:21:42.6&11.48&117.5&11235\\
ARP256S&00:18:50.90&-10:22:37.0&11.48&117.5&11235\\
MCG+12-02-001&00:54:04.20&+73:05:06.0&11.5&69.8&10169 \\
IC-1623E&01:07:46.49&-17:30:22.5&11.71&85.5&7219\\
IC-1623W&01:07:47.42&-17:30:25.9&11.71&85.5&7219\\
MCG-03-04-014&01:10:08.90&-16:51:10.0&11.65&144&11235\\
CGCG436-030&01:20:02.70&+14:21:43.0&11.69&134&11235\\
2MASXJ01385289-1027113&01:38:52.90&-10:27:11.0&11.85&198&11235\\
IIIZW035&01:44:30.52&+17:06:07.9&11.64&119&11235\\
NGC0695&01:51:14.20&+22:34:57.0&11.68&139&11235\\
MRK1034W&02:23:18.91&+32:11:19.2&11.64&145&11235\\
MRK1034E&02:23:22.00&+32:11:50.0&11.64&145&11235\\
UGC02369S&02:54:01.77&+14:58:14.2&11.67&136&11235\\
UGC02369N&02:54:01.81&+14:58:35.6&11.67&136&11235\\
IRASF03359+1523&03:38:46.95&+15:32:54.5&11.55&152&11235\\
ESO550-IG025N&04:21:19.87&-18:48:36.9&11.51&138.5&11235\\
ESO550-IG025S&04:21:20.06&-18:48:56.3&11.51&138.5&11235\\
NGC1614&04:33:59.91&-08:34:44.3&11.65&67.8&9726\\
ESO203-IG001&04:46:49.42&-48:33:31.1&11.86&235&11235\\
VII-Zw-031&05:16:46.60&+79:40:13.0&11.99&240&9726 \\
ESO255-IG007N&06:27:22.04&-47:10:42.0&11.9&173&11235\\
ESO255-IG007S&06:27:23.12&-47:11:03.1&11.9&173&11235\\
AM0702-601N&07:03:24.10&-60:15:23.0&11.64&141&11235\\
AM0702-601S&07:03:28.63&-60:16:42.9&11.64&141&11235\\
2MASXJ08370182-4954302&08:37:01.80&-49:54:30.0&11.62&210&11235\\
NGC2623&08:38:24.00&+25:45:17.0&11.6&84.1&7219\\
ESO060-IG016&08:52:30.77&-69:01:59.8&11.82&210&11235\\
IRAS-F08572+3915&09:00:25.40&+39:03:54.0&12.16&264&9726 \\
UGC05101&09:35:51.40&+61:21:11.0&12.01&177&9726\\
NGC3256&10:27:51.03&-43:54:18.2&11.64&38.9&9735\\
IRAS-F10565+2448&10:59:18.20&+24:32:37.0&12.08&197&7219 \\
IRAS-F11231+1456&11:25:45.00&+14:40:36.0&11.64&157&9726 \\
NGC3690W&11:28:30.91&+58:33:45.2&11.93&50.7&9726\\
NGC3690E&11:28:33.50&+58:33:45.2&11.93&50.7&9726\\
IRAS-F12112+0305&12:13:45.90&+02:48:39.0&12.36&340&7219\\
WKK0787&12:14:22.10&-56:32:33.0&11.65&114.5&11235\\
WKK2031&13:15:06.30&-55:09:23.0&12.32&144&11235\\
UGC08335W&13:15:31.15&+62:07:44.4&11.81&142&11235\\
UGC08335E&13:15:35.29&+62:07:27.5&11.81&142&11235\\
UGC08387&13:20:35.30&+34:08:22.0&11.73&110&7219 \\
NGC5256&13:38:17.79&+48:16:35.1&11.56&129&7328\\
NGC5257&13:39:52.97&+00:50:22.5&11.56&129&11235\\
UGC08696&13:44:41.90&+55:53:12.1&12.21&173&9726\\
NGC5331S&13:52:16.17&+02:06:01.2&11.66&155&11235\\
NGC5331N&13:52:16.54&+02:06:28.8&11.66&155&11235\\
IRAS-F14348-1447&14:37:38.20&-15:00:24.0&12.39&387&7219\\
IRAS-F14378-3651&14:40:58.91&-37:04:31.8&12.23&315&7896 \\
UGC09618S&14:57:00.34&+24:36:24.7&11.74&157&11235\\
UGC09618N&14:57:00.70&+24:37:01.5&11.74&157&11235\\
ESO099-G004&15:24:58.20&-63:07:34.0&11.74&137&11235\\
IRAS-F15250+3608&15:26:59.30&+35:58:37.0&12.08&254&7219\\
UGC09913&15:34:57.20&+23:30:10.9&12.28&87.9&9726\\
NGC6090&16:11:40.60&+52:27:25.0&11.58&137&7219 \\
2MASXJ16191179-0754026&16:19:11.80&-07:54:03.0&11.62&128&11235\\
ESO069-IG006N&16:38:11.84&-68:26:10.4&11.98&212&11235\\
ESO069-IG006S&16:38:13.48&-68:27:19.3&11.98&212&11235\\
IRAS16399-0937&16:42:40.20&-09:43:14.0&11.63&114&11235\\
NGC6240&16:52:58.70&+02:24:04.0&11.93&116&7219\\
IRASF17132+5313&17:14:20.24&+53:10:30.8&11.96&232&11235\\
IRAS-F17138-1017&17:16:35.60&-10:20:38.0&11.49&84&10169\\
IRAS-F17207-0014&17:23:22.20&-00:17:02.0&12.46&198&7219\\
IRAS18090+0130&18:11:38.42&+01:31:38.8&11.65&134&11235\\
IC4689&18:13:40.28&-57:44:53.5&11.62&81.9&11235\\
IRAS18293-3413&18:32:41.22&-34:11:26.7&11.88&86&11235\\
NGC6670W&18:33:33.77&+59:53:15.6&11.65&129.5&11235\\
NGC6670E&18:33:37.70&+59:53:23.0&11.65&129.5&11235\\
NGC6786S&19:10:53.90&+73:24:37.0&11.49&113&11235\\
UGC11415&19:10:53.90&+73:24:37.7&11.49&113&11235\\
NGC6786N&19:11:04.44&+73:25:37.4&11.49&113&11235\\
ESO593-IG008&19:14:30.98&-21:19:06.8&11.93&222&11235\\
IRAS-F19297-0406&19:32:22.28&-04:00:01.1&12.45&395&7896\\
IRAS19542+1110&19:56:35.39&+11:19:03.0&12.12&295&11235\\
IRAS20351+2521&20:37:17.80&+25:31:38.0&11.61&151&11235\\
IIZW096&20:57:23.94&+17:07:39.5&11.94&161&11235\\
ESO286-IG019&20:58:26.80&-42:39:00.0&12.06&193&11235\\
IRAS21101+5810&21:11:29.82&+58:23:07.2&11.81&174&11235\\
ESO239-IG002&22:49:39.90&-48:50:58.0&11.84&191&11235\\
IRAS-F22491-1808&22:51:49.30&-17:52:24.0&12.2&351&7219\\
NGC7469  / IC5283&23:03:17.88&+08:53:39.3&11.65&70.8&11235\\
ESO148-IG002&23:15:46.79&-59:03:13.0&12.06&199&7896\\
IC5298&23:16:00.71&+25:33:24.0&11.6&119&11235\\
ESO077-IG014&23:21:04.44&-69:12:54.8&11.76&186&11235\\
NGC7674&23:27:56.70&+08:46:45.0&11.56&125&11235\\
IRASF23365+3604&23:39:01.30&+36:21:09.8&12.2&287&11235\\
IRAS23436+5257&23:46:05.59&+53:14:01.0&11.57&149&11235\\
UGC12812W / MRK0331&23:51:18.73&+20:34:42.9&11.5&79.3&11235\\
UGC12812E  / MRK0331&23:51:26.80&+20:35:10.0&11.5&79.3&11235\\
\enddata
\tablecomments{Summary of the properties of our 73 LIRG systems (88 pointings). Column (1): Source Name from NED, Column (2): right ascension (J2000), Column (3): source declination (J2000), Column (4): The luminosity distance in Mpc (adopting $H_0=70$km~s$^{-1}$~Mpc), as provided by NED, Column (5): The total infrared luminosity in log$_{10}$ Solar units, Column (6): The data origin given by the ID of the observational program.}
\label{tab:obs}
\end{deluxetable}

\begin{deluxetable}{lccclllllll}
\tabletypesize{\tiny}
\rotate
\tablewidth{0pt}
\tablecaption{Bulge Parameters and Merger Classification}
\tablehead{\colhead{Name}  & \colhead{RA$_{Bulge}$} & \colhead{DEC$_{Bulge}$} &  \colhead{M. Class.} & \colhead{$L_{B}$} & \colhead{$R_{B}$} & \colhead{$n_{B}$} & \colhead{$M_{BH}$} & \colhead{L$_{B}$/R$_{B}^2$} &  \colhead{$L_{PSF}$} & \colhead{GALFIT comp.} \\
       & (J2000) & (J2000) &  &  log10[L$_\sun$] & [kpc] & & log10[$M_\sun$] & [$10^{10}\times$L$_{\sun}$/kpc$^2$]& log10[L$_\sun$] & } 
\startdata
NGC0034 & 00:11:06.544 & -12:06:27.24 & 5& 11.12 $\pm$ 10.08 & 0.13 $\pm$ 0.0 & 1.87 $\pm$ 0.01 & 8.56 $\pm$ 0.09 & 777.34 & 8.97 & 2\\ 
ARP256S & 00:18:50.870 & -10:22:36.46 & 3& 10.87 $\pm$ 9.54 & 0.35 $\pm$ 0.03 & 2.62 $\pm$ 0.16 & 8.27 $\pm$ 0.07 & 59.76 & 9.35 & 1\\ 
MCG-03-04-014 & 01:10:08.912 & -16:51:09.58 & 0& 11.32 $\pm$ 9.98 & 1.05 $\pm$ 0.04 & 2.07 $\pm$ 0.04 & 8.79 $\pm$ 0.1 & 18.85 & 9.44 & 2\\ 
CGCG436-030 & 01:20:02.632 & +14:21:42.26 & 2& 10.84 $\pm$ 9.4 & 0.22 $\pm$ 0.02 & 6.76 $\pm$ 0.42 & 8.23 $\pm$ 0.07 & 136.95 & 9.46 & 2\\ 
2MASXJ01385289-1027113 & 01:38:52.857 & -10:27:11.38 & 5& 10.42 $\pm$ 8.39 & 0.38 $\pm$ 0.0 & 1.39 $\pm$ 0.02 & 7.75 $\pm$ 0.08 & 18.72 & 8.66 & 2\\ 
IIIZW035 & 01:44:30.546 & +17:06:09.04 & 3& 11.15 $\pm$ 9.11 & 2.07 $\pm$ 0.02 & 2.32 $\pm$ 0.04 & 8.59 $\pm$ 0.08 & 3.29 & 8.66 & 2\\ 
NGC0695 & 01:51:14.333 & +22:34:56.02 & 0& 11.41 $\pm$ 10.07 & 1.85 $\pm$ 0.11 & 3.73 $\pm$ 0.07 & 8.9 $\pm$ 0.1 & 7.54 & 8.91 & 2\\ 
MRK1034W & 02:23:18.953 & +32:11:18.69 & 2& 11.1 $\pm$ 9.54 & 2.08 $\pm$ 0.1 & 7.75 $\pm$ 0.11 & 8.54 $\pm$ 0.08 & 2.93 & 8.82 & 3\\ 
MRK1034E & 02:23:21.949 & +32:11:48.83 & 2& 11.56 $\pm$ 9.52 & 2.91 $\pm$ 0.02 & 1.14 $\pm$ 0.01 & 9.07 $\pm$ 0.12 & 4.29 & - & 3\\ 
IRASF03359+1523 & 03:38:46.351 & +15:32:54.69 & 3& 10.99 $\pm$ 8.96 & 2.82 $\pm$ 0.07 & 7.45 $\pm$ 0.07 & 8.41 $\pm$ 0.07 & 1.24 & - & 2\\ 
ESO550-IG02N & 04:21:19.977 & -18:48:39.39 & 2& 10.59 $\pm$ 8.55 & 0.19 $\pm$ 0.0 & 1.37 $\pm$ 0.01 & 7.94 $\pm$ 0.07 & 107.86 & 8.88 & 2\\ 
ESO550-IG02S & 04:21:20.017 & -18:48:57.17 & 2& 10.5 $\pm$ 9.46 & 0.2 $\pm$ 0.0 & 1.29 $\pm$ 0.01 & 7.84 $\pm$ 0.09 & 81.35 & - & 2\\ 
NGC1614 & 04:34:00.015 & -08:34:45.13 & 5& 10.9 $\pm$ 8.86 & 0.08 $\pm$ 0.0 & 1.55 $\pm$ 0.02 & 8.31 $\pm$ 0.07 & 1309.26 & 10.23 & 3\\ 
ESO203-IG001 & 04:46:48.986 & -48:33:35.10 & 3& 10.45 $\pm$ 8.42 & 2.41 $\pm$ 0.05 & 2.94 $\pm$ 0.04 & 7.79 $\pm$ 0.08 & 0.49 & - & 1\\ 
ESO203-IG001 & 04:46:49.537 & -48:33:29.90 & 3& 10.87 $\pm$ 8.84 & 1.37 $\pm$ 0.02 & 2.02 $\pm$ 0.01 & 8.27 $\pm$ 0.07 & 3.93 & 8.3 & 2\\ 
VII-Zw-031 & 05:16:46.502 & +79:40:12.85 & 0& 11.34 $\pm$ 10.38 & 1.09 $\pm$ 0.18 & 4.47 $\pm$ 0.28 & 8.81 $\pm$ 0.11 & 18.35 & - & 2\\ 
ESO255-IG007N & 06:27:21.754 & -47:10:35.78 & 3& 11.25 $\pm$ 10.36 & 0.61 $\pm$ 0.1 & 1.35 $\pm$ 0.2 & 8.71 $\pm$ 0.11 & 47.82 & - & 1\\ 
ESO255-IG007N & 06:27:22.587 & -47:10:46.79 & 3& 10.84 $\pm$ 9.76 & 0.96 $\pm$ 0.09 & 1.6 $\pm$ 0.13 & 8.24 $\pm$ 0.08 & 7.57 & - & 1\\ 
ESO255-IG007S & 06:27:23.090 & -47:11:02.73 & 3& 11.11 $\pm$ 9.38 & 9.37 $\pm$ 0.6 & 3.12 $\pm$ 0.1 & 8.55 $\pm$ 0.08 & 0.15 & - & 2\\ 
AM0702-601N & 07:03:24.257 & -60:15:22.53 & 1& 10.71 $\pm$ 9.93 & 1.81 $\pm$ 0.38 & 4.45 $\pm$ 0.39 & 8.09 $\pm$ 0.11 & 1.58 & - & 3\\ 
AM0702-601S & 07:03:28.627 & -60:16:44.67 & 1& 11.43 $\pm$ 10.4 & 2.14 $\pm$ 0.03 & 3.14 $\pm$ 0.04 & 8.92 $\pm$ 0.11 & 5.92 & - & 2\\ 
NGC2623 & 08:38:24.094 & +25:45:16.75 & 5& 10.61 $\pm$ 8.57 & 0.13 $\pm$ 0.0 & 3.39 $\pm$ 0.04 & 7.97 $\pm$ 0.07 & 225.78 & 8.45 & 2\\ 
ESO060-IG016 & 08:52:32.131 & -69:01:55.50 & 3& 10.84 $\pm$ 9.88 & 1.27 $\pm$ 0.19 & 2.04 $\pm$ 0.22 & 8.24 $\pm$ 0.09 & 4.33 & - & 1\\ 
IRAS-F08572+3915 & 09:00:25.593 & +39:03:50.60 & 3& 9.92 $\pm$ 7.88 & 0.44 $\pm$ 0.0 & 0.91 $\pm$ 0.02 & 7.16 $\pm$ 0.13 & 4.18 & - & 2\\ 
UGC05101 & 09:35:51.629 & +61:21:11.89 & 5& 11.48 $\pm$ 9.75 & 0.84 $\pm$ 0.02 & 3.05 $\pm$ 0.07 & 8.98 $\pm$ 0.11 & 43.14 & 10.12 & 2\\ 
IRAS-F10565+2448 & 10:59:18.146 & +24:32:34.54 & 2& 11.65 $\pm$ 10.61 & 1.0 $\pm$ 0.01 & 2.73 $\pm$ 0.02 & 9.18 $\pm$ 0.13 & 44.7 & 10.09 & 1\\ 
IRAS-F11231+1456 & 11:25:45.077 & +14:40:36.07 & 1& 11.4 $\pm$ 9.84 & 3.29 $\pm$ 0.15 & 5.06 $\pm$ 0.08 & 8.88 $\pm$ 0.1 & 2.31 & - & 2\\ 
NGC3690E & 11:28:33.690 & +58:33:46.35 & 3& 10.85 $\pm$ 9.82 & 0.68 $\pm$ 0.07 & 3.54 $\pm$ 0.13 & 8.25 $\pm$ 0.08 & 15.57 & - & 2\\ 
IRAS-F12112+0305 & 12:13:45.904 & +02:48:39.26 & 4& 11.0 $\pm$ 9.44 & 0.32 $\pm$ 0.02 & 5.98 $\pm$ 0.4 & 8.42 $\pm$ 0.08 & 99.63 & 8.91 & 1\\ 
WKK0787 & 12:14:22.098 & -56:32:33.26 & 0& 11.43 $\pm$ 10.4 & 1.44 $\pm$ 0.01 & 2.95 $\pm$ 0.02 & 8.93 $\pm$ 0.11 & 13.07 & 9.6 & 1\\ 
WKK2031 & 13:15:06.339 & -55:09:22.72 & 5& 11.56 $\pm$ 9.22 & 0.46 $\pm$ 0.0 & 1.11 $\pm$ 0.01 & 9.07 $\pm$ 0.11 & 170.09 & 10.43 & 2\\ 
UGC08335W & 13:15:30.799 & +62:07:45.31 & 2& 10.43 $\pm$ 9.3 & 0.49 $\pm$ 0.06 & 4.72 $\pm$ 0.34 & 7.76 $\pm$ 0.09 & 11.31 & - & 2\\ 
UGC08335E & 13:15:35.023 & +62:07:28.86 & 2& 11.23 $\pm$ 9.19 & 0.76 $\pm$ 0.01 & 3.72 $\pm$ 0.04 & 8.69 $\pm$ 0.09 & 29.4 & 8.44 & 2\\ 
UGC08387 & 13:20:35.319 & +34:08:22.39 & 4& 10.51 $\pm$ 8.47 & 0.55 $\pm$ 0.01 & 1.27 $\pm$ 0.02 & 7.85 $\pm$ 0.08 & 10.55 & - & 2\\ 
NGC5256 & 13:38:17.270 & +48:16:32.10 & 3& 10.87 $\pm$ 8.84 & 0.62 $\pm$ 0.0 & 1.02 $\pm$ 0.01 & 8.28 $\pm$ 0.07 & 19.45 & - & 2\\ 
NGC5256 & 13:38:17.764 & +48:16:41.19 & 3& 10.96 $\pm$ 9.77 & 0.76 $\pm$ 0.07 & 3.47 $\pm$ 0.12 & 8.37 $\pm$ 0.08 & 15.65 & 8.39 & 2\\ 
NGC5257 & 13:39:53.767 & +00:50:26.78 & 2& 11.03 $\pm$ 9.6 & 2.29 $\pm$ 0.16 & 6.22 $\pm$ 0.16 & 8.46 $\pm$ 0.08 & 2.06 & - & 2\\ 
IRAS-F14348-1447 & 14:37:38.285 & -15:00:24.05 & 4& 10.92 $\pm$ 9.48 & 0.36 $\pm$ 0.01 & 1.51 $\pm$ 0.06 & 8.32 $\pm$ 0.07 & 63.44 & 9.64 & 2\\ 
IRAS-F14348-1447 & 14:37:38.402 & -15:00:21.13 & 4& 11.57 $\pm$ 10.13 & 5.87 $\pm$ 0.36 & 4.7 $\pm$ 0.12 & 9.08 $\pm$ 0.12 & 1.08 & 8.74 & 1\\ 
IRAS-F14378-3651 & 14:40:59.013 & -37:04:31.89 & 6& 11.19 $\pm$ 10.16 & 0.24 $\pm$ 0.0 & 1.43 $\pm$ 0.01 & 8.65 $\pm$ 0.1 & 262.25 & - & 2\\ 
UGC09618S & 14:57:00.327 & +24:36:24.02 & 1& 11.56 $\pm$ 9.53 & 7.23 $\pm$ 0.05 & 3.39 $\pm$ 0.01 & 9.08 $\pm$ 0.12 & 0.7 & 8.95 & 2\\ 
ESO099-G004 & 15:24:57.954 & -63:07:29.67 & 3& 10.92 $\pm$ 8.89 & 0.4 $\pm$ 0.01 & 5.42 $\pm$ 0.16 & 8.33 $\pm$ 0.07 & 51.13 & - & 1\\ 
IRAS-F15250+3608 & 15:26:59.423 & +35:58:37.22 & 5& 10.85 $\pm$ 9.86 & 0.56 $\pm$ 0.08 & 3.21 $\pm$ 0.24 & 8.25 $\pm$ 0.09 & 22.81 & - & 2\\ 
NGC6090 & 16:11:40.918 & +52:27:27.22 & 4& 10.07 $\pm$ 8.64 & 0.16 $\pm$ 0.01 & 2.16 $\pm$ 0.12 & 7.34 $\pm$ 0.11 & 46.12 & - & 2\\ 
2MASXJ16191179-0754026 & 16:19:11.787 & -07:54:03.02 & 5& 11.33 $\pm$ 9.6 & 4.98 $\pm$ 0.17 & 4.09 $\pm$ 0.08 & 8.81 $\pm$ 0.1 & 0.86 & - & 2\\ 
ESO069-IG006S & 16:38:13.472 & -68:27:16.83 & 2& 11.25 $\pm$ 9.52 & 1.31 $\pm$ 0.03 & 3.24 $\pm$ 0.03 & 8.72 $\pm$ 0.09 & 10.48 & - & 2\\ 
IRAS16399-0937 & 16:42:40.141 & -09:43:13.20 & 3& 10.71 $\pm$ 9.16 & 1.03 $\pm$ 0.02 & 1.26 $\pm$ 0.07 & 8.09 $\pm$ 0.07 & 4.91 & 8.69 & 2\\ 
IRAS16399-0937 & 16:42:40.178 & -09:43:18.74 & 3& 10.63 $\pm$ 9.55 & 0.51 $\pm$ 0.05 & 2.93 $\pm$ 0.15 & 7.99 $\pm$ 0.08 & 16.38 & - & 2\\ 
NGC6240 & 16:52:58.886 & +02:24:03.20 & 4& 11.29 $\pm$ 10.25 & 0.23 $\pm$ 0.0 & 1.6 $\pm$ 0.01 & 8.76 $\pm$ 0.1 & 376.66 & 9.45 & 1\\ 
NGC6240 & 16:52:58.934 & +02:24:04.92 & 4& 10.81 $\pm$ 9.38 & 0.2 $\pm$ 0.01 & 1.82 $\pm$ 0.07 & 8.2 $\pm$ 0.07 & 155.08 & - & 2\\ 
IRASF17132+5313 & 17:14:19.802 & +53:10:28.89 & 2& 11.09 $\pm$ 9.36 & 0.71 $\pm$ 0.02 & 5.41 $\pm$ 0.08 & 8.53 $\pm$ 0.08 & 24.46 & 9.01 & 2\\ 
IRASF17132+5313 & 17:14:20.453 & +53:10:31.99 & 2& 11.04 $\pm$ 10.4 & 1.16 $\pm$ 0.31 & 2.81 $\pm$ 0.32 & 8.47 $\pm$ 0.14 & 8.12 & - & 3\\ 
IRAS-F17207-0014 & 17:23:21.953 & -00:17:00.75 & 5& 10.39 $\pm$ 9.06 & 0.33 $\pm$ 0.01 & 0.96 $\pm$ 0.05 & 7.72 $\pm$ 0.09 & 23.41 & - & 2\\ 
IRAS18090+0130 & 18:11:38.412 & +01:31:39.99 & 2& 10.54 $\pm$ 8.5 & 0.26 $\pm$ 0.0 & 1.12 $\pm$ 0.03 & 7.89 $\pm$ 0.08 & 51.38 & - & 2\\ 
IC4689 & 18:13:40.386 & -57:44:54.18 & 2& 10.32 $\pm$ 8.29 & 0.22 $\pm$ 0.0 & 2.86 $\pm$ 0.03 & 7.64 $\pm$ 0.09 & 45.48 & 8.15 & 2\\ 
IRAS18293-3413 & 18:32:41.139 & -34:11:27.61 & 1& 11.58 $\pm$ 9.84 & 1.79 $\pm$ 0.08 & 5.67 $\pm$ 0.13 & 9.09 $\pm$ 0.12 & 11.86 & - & 2\\ 
NGC6670W & 18:33:33.972 & +59:53:17.49 & 2& 11.09 $\pm$ 9.05 & 2.04 $\pm$ 0.02 & 1.78 $\pm$ 0.02 & 8.52 $\pm$ 0.08 & 2.94 & - & 2\\ 
NGC6670E & 18:33:37.721 & +59:53:23.07 & 2& 10.65 $\pm$ 9.62 & 0.88 $\pm$ 0.0 & 0.62 $\pm$ 0.01 & 8.02 $\pm$ 0.09 & 5.88 & - & 2\\ 
NGC6786N & 19:11:04.302 & +73:25:33.42 & 2& 10.59 $\pm$ 9.56 & 0.1 $\pm$ 0.0 & 2.87 $\pm$ 0.05 & 7.95 $\pm$ 0.09 & 398.67 & 9.14 & 2\\ 
ESO593-IG008 & 19:14:31.099 & -21:19:09.33 & 4& 11.35 $\pm$ 10.09 & 1.28 $\pm$ 0.1 & 2.37 $\pm$ 0.14 & 8.83 $\pm$ 0.1 & 13.61 & 7.49 & 1\\ 
IRAS19542+1110 & 19:56:35.787 & +11:19:05.09 & 0& 11.5 $\pm$ 10.17 & 2.04 $\pm$ 0.06 & 2.95 $\pm$ 0.11 & 9.01 $\pm$ 0.11 & 7.7 & 11.56 & 2\\ 
IIZW096 & 20:57:24.475 & +17:07:39.89 & 3& 10.09 $\pm$ 9.38 & 0.59 $\pm$ 0.16 & 3.58 $\pm$ 0.44 & 7.37 $\pm$ 0.15 & 3.59 & - & 2\\ 
ESO286-IG019 & 20:58:26.801 & -42:39:00.20 & 5& 11.18 $\pm$ 9.15 & 0.7 $\pm$ 0.01 & 1.92 $\pm$ 0.02 & 8.64 $\pm$ 0.08 & 31.2 & 9.04 & 2\\ 
IRAS21101+5810 & 21:11:30.392 & +58:23:03.38 & 2& 10.42 $\pm$ 8.69 & 0.39 $\pm$ 0.01 & 2.16 $\pm$ 0.02 & 7.75 $\pm$ 0.08 & 17.5 & - & 2\\ 
ESO239-IG002 & 22:49:39.889 & -48:50:58.20 & 5& 11.18 $\pm$ 10.15 & 0.16 $\pm$ 0.0 & 3.02 $\pm$ 0.02 & 8.63 $\pm$ 0.1 & 571.59 & 9.96 & 2\\ 
IRAS-F22491-1808 & 22:51:49.354 & -17:52:24.08 & 4& 10.14 $\pm$ 8.8 & 0.32 $\pm$ 0.02 & 0.97 $\pm$ 0.15 & 7.42 $\pm$ 0.11 & 13.39 & - & 1\\ 
NGC7469  / IC5283 & 23:03:17.964 & +08:53:36.72 & 2& 9.78 $\pm$ 8.04 & 0.24 $\pm$ 0.01 & 1.92 $\pm$ 0.05 & 7.0 $\pm$ 0.14 & 10.42 & - & 3\\ 
IC5298 & 23:16:00.682 & +25:33:24.02 & 0& 11.16 $\pm$ 9.42 & 1.17 $\pm$ 0.01 & 1.17 $\pm$ 0.03 & 8.6 $\pm$ 0.08 & 10.41 & 9.63 & 3\\ 
ESO077-IG014 & 23:21:03.620 & -69:13:00.99 & 2& 11.32 $\pm$ 9.28 & 3.44 $\pm$ 0.03 & 2.33 $\pm$ 0.02 & 8.79 $\pm$ 0.09 & 1.75 & - & 3\\ 
ESO077-IG014 & 23:21:05.329 & -69:12:47.30 & 2& 11.47 $\pm$ 9.91 & 2.55 $\pm$ 0.11 & 4.68 $\pm$ 0.08 & 8.96 $\pm$ 0.11 & 4.52 & 9.25 & 3\\ 
NGC7674 & 23:27:56.717 & +08:46:44.40 & 2& 11.07 $\pm$ 10.03 & 0.37 $\pm$ 0.0 & 4.27 $\pm$ 0.04 & 8.5 $\pm$ 0.09 & 86.46 & 10.37 & 3\\ 
IRASF23365+3604 & 23:39:01.269 & +36:21:08.54 & 5& 10.96 $\pm$ 9.92 & 0.26 $\pm$ 0.0 & 1.65 $\pm$ 0.02 & 8.37 $\pm$ 0.09 & 134.53 & - & 2\\ 
IRAS23436+5257 & 23:46:05.512 & +53:14:01.40 & 4& 10.58 $\pm$ 8.85 & 0.24 $\pm$ 0.0 & 1.5 $\pm$ 0.03 & 7.94 $\pm$ 0.08 & 67.86 & 9.56 & 2\\ 
UGC12812W & 23:51:18.675 & +20:34:41.59 & 1& 9.57 $\pm$ 8.86 & 1.64 $\pm$ 0.17 & 1.5 $\pm$ 0.03 & 6.77 $\pm$ 0.19 & 0.14 & - & 2\\ 
UGC12812E & 23:51:26.741 & +20:35:10.41 & 1& 11.06 $\pm$ 9.5 & 2.3 $\pm$ 0.15 & 7.43 $\pm$ 0.19 & 8.5 $\pm$ 0.08 & 2.19 & - & 3\\ 
\enddata
\tablecomments{Column (1): Source Name from NED, Column (2): right ascension (J2000) of bulge center, Column (3): declination (J2000) of bulge center, Column (4): The merger classification as described in \S~\ref{subsec:res-GALFIT}, Column (5): The apparent bulge luminosity with fit error in mag , Column (6): The bulge radius with fit error in kpc, Column (7): The S\'{e}rsic index of the bulge radial profile with fit error, Column (8): The mass of the central black hole estimated from the bulge luminosity \citep[based on the relation of][]{Mar03}, Column (9): The bulge surface density L$_{Bulge}$/R$_{Bulge^2}$, Column (10): The luminosity of the central PSF, Column (11): The number of GALFIT components (excluding PSF and background).}
\label{tab:galfit}
\end{deluxetable}

\clearpage
\newpage

\begin{deluxetable}{llrrrrr}
%\rotate
%\tablewidth{0pt}
\tabletypesize{\tiny}
\tablecaption{Merger Stage Results}
\tablehead{\colhead{Bulge property}  & \colhead{Statistics} & \colhead{1} & \colhead{2} &  \colhead{3}& \colhead{4} & \colhead{5/6}}
\startdata
lL log[L$_{\odot}$] &Mean & 11.38 & 11.09 & 10.88 & 11.05 & 11.14\\ 
 &Median & 11.41 & 11.04 & 10.87 & 10.87 & 11.12\\ 
 &Mean Error & 9.95 & 9.75 & 9.59 & 9.73 & 9.74\\ 
 &Std. Dev. & 9.87 & 9.97 & 9.72 & 9.8 & 9.71\\ 
R [kpc]  &Mean & 3.09 & 1.08 & 1.54 & 0.95 & 0.71\\ 
 &Median & 2.22 & 0.74 & 0.76 & 0.32 & 0.33\\ 
 &Mean Error & 0.14 & 0.04 & 0.09 & 0.05 & 0.02\\ 
 &Std. Dev. & 1.92 & 0.99 & 2.08 & 1.67 & 1.25\\ 
n$_s$ &Mean & 4.86 & 3.26 & 2.8 & 2.39 & 2.2\\ 
 &Median & 4.76 & 2.84 & 2.62 & 1.71 & 1.87\\ 
 &Mean Error & 0.14 & 0.09 & 0.12 & 0.11 & 0.05\\ 
 &Std. Dev. & 1.45 & 1.93 & 1.61 & 1.55 & 0.97\\ 
LSD log[$10^{10}\times$L$_{\sun}$/kpc$^2$] &Mean & 10.61 & 11.69 & 11.18 & 11.93 & 12.44\\ 
 &Median & 10.35 & 11.16 & 10.69 & 11.74 & 12.13\\ 
 &Mean Error & 9.69 & 10.64 & 10.43 & 10.89 & 11.2\\ 
 &Std. Dev. & 10.58 & 11.93 & 11.27 & 12.03 & 12.57\\ 
\enddata
\tablecomments{The mean, median, mean systematical error (from GALFIT), and the standard deviation of the bulge properties (NIR luminosity L, radius R, S\'{e}rsic index n$_s$, and luminosity surface density LSD) for all merger stages. Note that we have combined merger stage 5 and 6 as post-merger stage LIRGs.}
\label{tab:stages}
\end{deluxetable}

\begin{deluxetable}{llrrrrrrrrrr}
%\rotate
%\tablewidth{0pt}
\tabletypesize{\tiny}
\tablecaption{KS and MWU Test Results}
\tablehead{\colhead{Bulge property}  & \colhead{Test} & \colhead{1 - 2} & \colhead{1 - 3} &  \colhead{1 - 4)}& \colhead{1  -5/6} &  \colhead{2 - 3}&  \colhead{2 - 4} &  \colhead{2 - 5/6}&  \colhead{3 - 4} &  \colhead{3 - 5/6} &  \colhead{4 - 5/6} }
\startdata
L &K-S p & \underline{0.01} & \underline{0.01} & \underline{0.02} & \underline{0.04} & \underline{0.11} & 0.74 & 0.54 & 0.54 & \underline{0.11} & 0.74\\ 
&MWU p & \underline{0.01} & \underline{0.01} & \underline{0.01} & \underline{0.03} & 0.48 & 0.68 & 0.43 & 0.8 & \underline{0.09} & 0.42\\ 
R &K-S p & \underline{0.01} & \underline{0.01} & \underline{0.01} & \underline{0.01} & \underline{0.15} & 0.21 & \underline{0.12} & \underline{0.01} & \underline{0.01} & 0.88\\ 
&MWU p & \underline{0.01} & \underline{0.01} & \underline{0.01} & \underline{0.01} & 0.31 & 0.25 & \underline{0.09} & \underline{0.01} & \underline{0.01} & 0.85\\ 
n$_s$ &K-S p &  \underline{0.09} & \underline{0.05} & \underline{0.01} & \underline{0.01} & 0.54 & 0.19 & 0.32 & 0.42 & 0.36 & 0.96\\ 
&MWU p & 0.19 & \underline{0.05} & \underline{0.04} & \underline{0.01} & 0.55 & 0.19 & 0.17 & 0.38 & 0.41 & 0.95\\ 
LSD &K-S p & \underline{0.06} & 0.73 &\underline{ 0.01} & \underline{0.01} & 0.39 & 0.28 & \underline{0.04} & \underline{0.06} & \underline{0.01} & 0.37\\ 
&MWU p & \underline{0.06} & 0.76 & \underline{0.01} & \underline{0.01} & \underline{0.09} & 0.18 & \underline{0.02} & \underline{0.02} & \underline{0.01} & 0.21\\ 
\enddata
\tablecomments{The Kolmogorov-Smirnov (KS) and Mann-Whitney U (MWU) test results of the bulge properties (NIR luminosity L, radius R, S\'{e}rsic index n$_s$, and luminosity surface density LSD) for all combinations of merger stage populations. Note that we have combined merger stage 5 and 6 as post-merger stage LIRGs. The p values listed are the probability levels of the hypothesis that the two samples have the same distribution.  Large p values indicate that the two samples come from the same sample distribution while small p-values (p$\leq0.15$, underlined) are not drawn from the same parent population. The KS test tests differences in the shapes of the distributions of the two groups, not just the locations of the ranks as in the Mann-Whitney U test.}
\label{tab:ks_test}
\end{deluxetable}

\end{document}